\documentclass[10pt,journal,final,twocolumn]{IEEEtran}
\usepackage{epsfig,graphicx,subfigure,psfrag,amsmath,cases,bm}
\usepackage{latexsym,amssymb,algorithm,mathtools}
\usepackage{algorithmic}
\usepackage{color}
\usepackage{url}
\usepackage{scrtime}
\usepackage{stfloats}
\usepackage{tablefootnote}
\usepackage{cite}
\usepackage{amsfonts,mathabx}
\usepackage{textcomp}
\usepackage{amsmath,xcolor,capt-of}
\usepackage{multirow}
\usepackage{amsthm}
\usepackage{geometry}
\allowdisplaybreaks

\author{
\IEEEauthorblockN {Yiming Xu}
}

\newtheorem{remark}{Remark}
\newtheorem{lemma}{Lemma}

\newtheorem{T-Prob}{Transformed Problem}

\newcounter{TempEqCnt}

\DeclareMathOperator{\maxo}{maximize}
\DeclareMathOperator{\mino}{minimize}
\DeclareMathOperator{\diag}{\mathrm{diag}}

\DeclareMathOperator{\subto}{subject\hspace*{2mm}to}

\newcommand{\QED}{\hfill \ensuremath{\blacksquare}}

\newcommand{\myoverline}[1]{\overline{\overline{#1}}}

\begin{document}

\title{Sensing-assisted Robust SWIPT for Mobile Energy Harvesting Receivers in Networked ISAC Systems}

\author{
    Yiming Xu, \textit{Graduate Student Member, IEEE,}  Dongfang Xu, \textit{Member, IEEE,} \\ and Shenghui Song, \textit{Senior Member, IEEE} 
    \vspace*{-8mm}
\thanks{}
}

\maketitle

\begin{abstract}
Simultaneous wireless information
and power transfer (SWIPT) has been proposed to offer communication services and transfer power to the energy harvesting receiver (EHR) concurrently.
However, existing works mainly focused on static EHRs, without considering the location uncertainty caused by the movement of EHRs and location estimation errors.  
To tackle this issue, this paper considers the sensing-assisted SWIPT design in a networked integrated sensing and communication (ISAC) system in the presence of location uncertainty.
A two-phase robust design is proposed to reduce the location uncertainty and improve the power transfer efficiency. In particular, each time frame is divided into two phases, i.e., sensing and WPT phases, via time-splitting. The sensing phase performs collaborative sensing to localize the EHR, whose results are then utilized in the WPT phase for efficient WPT. 
To minimize the power consumption with given communication and power transfer requirements, a two-layer optimization framework is proposed to jointly optimize the time-splitting ratio, coordinated beamforming policy, and sensing node selection. Simulation results validate the effectiveness of the proposed design and demonstrate the existence of an optimal time-splitting ratio for given location uncertainty.
\end{abstract}


\section{Introduction}
Wireless power transfer (WPT) is a promising technique to prolong the lifetime of wireless devices in smart industrial Internet-of-things (IoT) networks \cite{9592187, 8214104, 7867832}. In particular, with WPT, energy harvesting receivers (EHRs) can harvest energy from the electromagnetic (EM) waves in radio frequency (RF) and obtain a stable and controllable energy supply. Furthermore, simultaneous wireless information and power transfer (SWIPT) was proposed to convey information and power for communication users (CUs) and EHRs concurrently. 
The core of the SWIPT system design is to allocate multiple types of resources like power, bandwidth, and time to meet the requirements of WPT and wireless information transfer (WIT) \cite{6489506}. To achieve efficient resource allocation, several schemes such as time switching \cite{6373669}, power splitting\cite{6567869}, and antenna switching\cite{6623062} were proposed to balance energy harvesting and information transmission performance.
\par
With the development of IoT networks, mobile sensors become more popular. In \cite{colmiais2024long}, Colmiais et al. designed a SWIPT system to power a moving mobile device attached to a walking individual. However, the investigation of SWIPT with mobile EHRs faces many new challenges and is still in its fancy.
In particular, it is more demanding to know the location information of mobile EHRs for accurate power transfer. One potential solution is to use sensing to localize the EHR and then perform information/energy transfer. To this end, the recently emerging integrated sensing and communication (ISAC) framework offers a promising solution. ISAC integrates sensing and communication services into one platform, allowing for the sharing of spectrum, hardware, and signal-processing modules \cite{9737357, 10086626, 9933849}. The sensing capability of ISAC has been utilized to support various intelligent applications such as Vehicle-to-Everything (V2X) networks, smart factories, and intelligent environment monitoring \cite{liu2020joint}. 
\par
The integration of ISAC with SWIPT has seen its applications.
In \cite{9977463, 10279127, 10296066}, the authors considered the integrated sensing, communication, and powering system.
Zeng et al. \cite{9977463} proposed a difference-of-convex-based algorithm to minimize the beampattern error while satisfying the requirements of the EHR and the CUs.
Chen et al. \cite{10279127} investigated the tradeoff between sensing, communication, and powering in the multi-functional
multiple-input multiple-output (MIMO) system by characterizing the achievable CRB-rate-energy region and its Pareto boundary. Yang et al. \cite{10296066} studied the intelligent reflecting surfaces (IRS)-assisted SWIPT service, where the transmit beamforming at the BS and the passive beamforming at the IRS were jointly optimized. However, existing works assumed static EHR and the location uncertainty caused by mobile EHR has not been tackled.
In particular, due to the movement of the EHR, the location estimation error is unavoidable, which necessitates a robust design to satisfy the energy harvesting requirement of the EHR.
\par
\par
\par
\par
In this paper, we consider the sensing-aided SWIPT for mobile EHRs in networked ISAC systems. Motivated by the BS cooperation in wireless communication networks, such as cloud radio access
networks (C-RAN) \cite{pana20225g}, cell-free MIMO systems \cite{chen2022survey}, and collaborative sensing \cite{xie2023collaborative}, the networked ISAC has attracted a lot of research interests \cite{xu2023joint, 10380513,10000730, liu2023joint}. Networked ISAC enables the collaborative sensing of the EHR from different perspectives, where the diversity gain helps improve the sensing performance. In addition, the self-interference problem can be avoided by separating the sensing transmitter and receiver in networked ISAC. However, networked ISAC also faces new challenges. For example, coordinated beamforming should be carefully designed for interference management purposes \cite{10207991}. In addition, the sensing nodes must be properly selected to serve as sensing transmitters or receivers \cite{xu2023joint}.
\par
Due to the movement of the EHR, the system needs to sense the EHRs and update their location information constantly. To achieve robust WPT for mobile EHRs, we propose a two-phase design, where each time frame is divided into two phases. In the first phase (sensing phase), the system performs collaborative sensing to obtain accurate location estimation. The sensing results are then utilized in the second phase (WPT phase) to transfer energy to EHRs. The time-splitting between the two phases is optimized to minimize the power consumption while guaranteeing the communication and power transfer requirements. 
\subsection{Challenges}
The objective of the sensing phase is to minimize the CRB of the location estimation with initial location uncertainty
while guaranteeing the communication requirement of the
CUs. This is achieved by jointly optimizing the beamforming policy and the sensing node
selection, which involves a robust CRB minimization problem where the location uncertainty is contained in the complex CRB matrix. Additionally, the sensing node selection is formulated as a challenging binary optimization problem. Moreover, the binary variables are coupled with the beamforming vectors which makes the problem very intractable. Based on the results in the sensing phase, the WPT phase aims to minimize the average power consumption while guaranteeing the requirements of both EHRs and the CUs.
The problem manifests itself as the robust WPT design with non-linear energy harvesting models, which is difficult to be efficiently solved.
Furthermore, the time-splitting ratio variable exists in both problems.
To this end, we develop a two-layer nested-loop optimization framework.  
\par
\subsection{Contributions}
The main contributions of this paper are listed as follows.
\begin{itemize}
    \item We consider exploiting sensing to assist the SWIPT service for mobile EHRs in the networked ISAC system. A two-phase algorithm is designed to minimize the power consumption of the system. The two-phase design is formulated into two sub-problems, i.e., the sensing and WPT sub-problem, and solved by a nested-loop framework. In particular, we use line search to determine the time-splitting strategy in the outer layer. For each given time-splitting strategy, the two sub-problems are solved in order in the inner layer.
    \item
    To deal with the robust CRB minimization problem in the sensing phase, the generalized Petersen’s sign-definiteness lemma is exploited. An optimization framework combining the big-M, penalty method, and successive convex approximation (SCA) technique is proposed to solve the non-convex binary constraint and the variable coupling problem. 
    In the WPT phase, fractional programming is utilized to deal with the fractional relation between the input power and the received power at the EHR. A series of transformations including S-procedure are used to handle the robust WPT with location uncertainty.
    \item The effectiveness of the proposed two-phase sensing-assisted SWIPT design is validated by simulation results. It is shown that the sensing phase can effectively improve WPT efficiency and save a lot of power. In addition, the proposed time-splitting design is robust to the initial location uncertainty. Moreover, for a given location uncertainty, there exists an optimal time-splitting ratio between the sensing and WPT phases.
\end{itemize}
\par
The rest of the paper is organized as follows. In Section II, we present the system model and the two-phase design. In section III, we define the performance metrics for sensing and communication and formulate the problem. A two-layer optimization framework is proposed in Section IV to solve the problem and numerical results are presented in Section V to validate the effectiveness of the proposed algorithm. Finally, Section VI concludes the paper.
\par
\textit{Notations:} 
Vectors and matrices are denoted by boldface lowercase and boldface capital letters, respectively. $\mathbb{R}^{M\times N}$ and $\mathbb{C}^{M\times N}$ represent the space of the $M\times N$ real-valued and complex-valued matrices, respectively. $|\cdot|$ and $||\cdot||_2$ denote the absolute value of a complex scalar and the $l_2$-norm of a vector, respectively. $(\cdot)^T$ and $(\cdot)^H$ stand for the transpose and the conjugate transpose of their arguments, respectively. $\mathbf{I}_{N}$ refers to the $N$ by $N$ identity matrix. $\mathrm{tr}(\mathbf{A})$ and $\mathrm{rank}(\mathbf{A})$ denote the trace and the rank of matrix $\mathbf{A}$, respectively. $\mathbf{A}\succeq\mathbf{0}$ indicates that $\mathbf{A}$ is a positive semidefinite matrix. $[\mathbf{A}]_{i,i}$ denotes the $i$-th diagonal element of matrix $\mathbf{A}$. $\Re\{\cdot\}$ and $\Im\{\cdot\}$ represent the real and imaginary parts of a complex number, respectively. Vectorization of matrix $\mathbf{A}$ is denoted by $\mathrm{vec}(\mathbf{A})$, and $\mathbf{A}\otimes\mathbf{B}$ represents the Kronecker product of two matrices $\mathbf{A}$ and $\mathbf{B}$. $\mathbb{E}[\cdot]$ refers to statistical expectation. $\overset{\Delta }{=}$ and $\sim$ stand for ``defined as'' and ``distributed as'', respectively. $\mathcal{O}(\cdot)$ is the big-O notation.
\begin{figure}[t]
\centering
\includegraphics[width=3.0in]{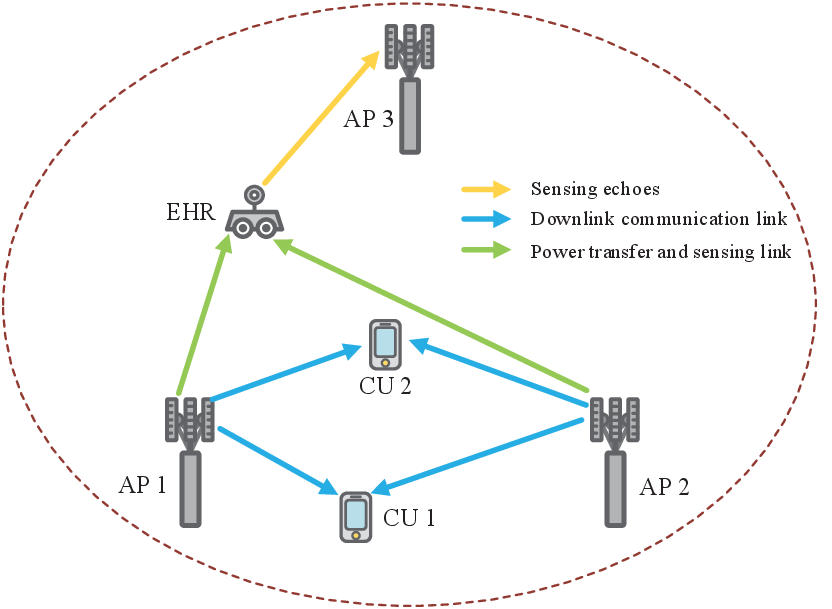}
\vspace*{-2mm}
\caption{Illustration of the considered networked ISAC system for SWIPT service.}
\label{figure:system_model}
\end{figure}
\begin{figure}[t]
\centering
\includegraphics[width=3.4in]{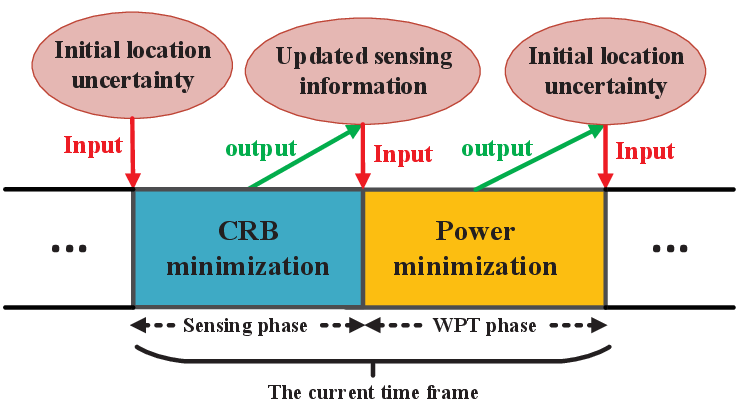}
\vspace*{-2mm}
\caption{Illustration of the proposed two-phase design.}
\label{figure:design_framework}
\end{figure}
\section{System Model}
As shown in Fig. \ref{figure:system_model}, we consider an ISAC-aided SWIPT system consisting of $Q$ access points (APs), $K$ communication users (CUs), and one EHR. APs are connected to a central controller for synchronization purposes \cite{9296833}. To focus on the investigation of SWIPT for the mobile EHR, we assume the EHR is a trustworthy device, such as a registered industrial robotic, and all APs are trustworthy for collaboration. This can be done by using the certificated APs to serve the reliable devices \cite{shaikh2016energy}. One typical use case of the considered model is in logistics systems. For example, to track the goods in a warehouse, one can attach wireless sensors to the goods. In such systems, the APs need to power the sensors that move together with the goods and provide communication services for other CUs at the same time.
\par
Each AP is equipped with a uniform linear array (ULA) comprising $N$ antennas. The CUs and the EHR are assumed to be single-antenna devices. For notation simplicity, we define sets $\mathcal{Q}\overset{\Delta }{=}\left\{1,\cdots,Q\right\}$ and $\mathcal{K}\overset{\Delta }{=}\left\{1,\cdots, K\right\}$ to collect the indices of the APs and CUs.
Assume a typical ISAC transmission frame has $L$ time-slots \cite{9652071} and the communication channels and the sensing parameters remain approximately constant within one frame \cite{10050406}.
To localize and transfer power to the EHR, we propose a two-phase design as illustrated in Fig. \ref{figure:design_framework}. 
In particular, each time frame is divided into a sensing phase of $\eta L$ time slots\footnote{$\eta L$ is rounded if it is not an integer.} and a WPT phase of $(1-\eta)L$ time slots with $\eta$ denoting the time-splitting ratio.
\par
Due to the movement of the EHR, unavoidable location uncertainty exists at the beginning of each time frame as shown in Fig. \ref{figure:design_framework}\footnote{The initial location error at each time frame can be modeled by the state prediction model in the tracking algorithms such as the extended Kalman filter. In this work, we focus on the design during one frame. The tracking algorithm can be easily applied for inter-frame tracking.}. In the first phase, collaborative sensing is performed to refine the location estimation. To avoid self-interference, we propose to select one AP as the sensing receiver to collect the echoes reflected from the EHR, while the other APs cooperatively transmit signals toward the CUs and the EHR.
Meanwhile, the EHR harvests energy from the ambient signals. 
In the second phase, WPT is performed based on the sensing result. 
In the meantime, the sensing receiver selected in the sensing phase continues to track the EHR to provide a stable localization performance for the following frame. Note that the requirements of the communication users need to be satisfied in both phases.
\par
\subsection{Signal Model}
To perform sensing and communication simultaneously, the $q$-th AP transmits the following DFRC signals at time slot $l \in \mathcal{L} \overset{\triangle}{=} \left[1,\cdots, L\right]$
\begin{equation}
\mathbf{x}_q[l] = a_q \left (\underset{k\in\mathcal{K}}{\sum} \mathbf{w}_{q,k}s_{k}[l] + \mathbf{r}_{q}[l]\right ),
\end{equation}
where $a_q$ denotes the AP selection variable. Specifically, $a_q=1$ means the $q$-th AP is selected as the transmitter while $a_q=0$ means it is selected as the sensing receiver. $\mathbf{w}_{q,k} \in \mathbb{C}^{N}$ is the beamforming vector at the $q$-th AP towards the $k$-th CU and $s_{k}[l]$ denotes the communication signal for the $k$-th CU at time slot $l$. 
Define the communication signal vector $\mathbf{s}[l] = \left[ s_1[l],\cdots, s_K[l]\right]^T \in \mathbb{C}^K$, whose entries are assumed to be independent and identically distributed (i.i.d.) Gaussian random variables with zero mean and unit variance, i.e., $\mathbb{E}\left[ \mathbf{s}[l] \mathbf{s}[l]^H \right] = \mathbf{I}_{K}$. $\mathbf{r}_q[l] \in \mathbb{C}^{N}$ represents the dedicated radar signal of the $q$-th AP at time slot $l$ and is independent with the communication signal, such that $ \mathbb{E}\left [ \mathbf{s}[l] \mathbf{r}_{q}^H[l] \right ] = \mathbf{0}_{K \times N}$.
Denote the covariance matrix of the radar signal as $\mathbf{R}_q \overset{\triangle}{=} \mathbb{E}\left [ \mathbf{r}_{q}[l] \mathbf{r}_{q}^H[l] \right ] $.
Then the covariance matrix of the transmitted signal $\mathbf{x}_q[l]$ is given by
\begin{eqnarray}
\mathbf{S}_{q} \overset{\triangle}{=} \mathbb{E} \left [ \mathbf{x}_q[l] \mathbf{x}_q[l]^H \right ] = a_q \left ( \sum_{k \in \mathcal{K}} \mathbf{w}_{q,k} \mathbf{w}_{q,k}^H + \mathbf{R}_q \right ).
\end{eqnarray}
\par
\vspace*{1mm}
\noindent  \textit{A.1: Received Communication Signal}
\vspace*{1mm}
\par
Denote $\mathbf{h}_{q,k} \in \mathbb{C}^{N}$ as the channel vector between the $q$-th AP and the $k$-th CU.
For simplicity, we define the overall channel vector, beamforming vector, and dedicated radar signal vector as $\mathbf{h}_k \overset{\triangle}{=} [\mathbf{h}^T_{1,k},\cdots,\mathbf{h}^T_{Q,k}]^T$, $\mathbf{w}_k \overset{\triangle}{=} [\mathbf{w}^T_{1,k},\cdots,\mathbf{w}^T_{Q,k}]^T$, and $\mathbf{r}[l] \overset{\triangle}{=} \big[\mathbf{r}^T_{1}[l],\cdots,\mathbf{r}^T_{Q}[l]\big]^T$, respectively. 
Furthermore, we define the association matrix $\mathbf{A}_q = \mathrm{diag}\{\mathbf{1}_q\}\otimes \mathbf{I}_N$, where $\mathbf{1}_q$ is a vector of length $Q$ whose $q$-th element is $0$ while other elements are $1$. Then the received signal $y_k[l]$ of the $k$-th CU at time slot $l$ can be expressed as
\begin{eqnarray} \label{comm_signal_2}
y_{k}[l] &\hspace*{-2mm}=\hspace*{-2mm}& \underbrace{\mathbf{h}_k^H \mathbf{A}_q \mathbf{w}_k s_k[l]}_{\text{Desired information signal}} + \underbrace{\underset{k' \in \mathcal{K} \setminus \{k\} }{\sum} \mathbf{h}_k^H \mathbf{A}_q \mathbf{w}_{k'} s_{k'}[l]}_{\text{Interference from dedicated radar signals}} \notag \\
&\hspace*{-2mm} + \hspace*{-2mm}& \underbrace{\mathbf{h}_k^H \mathbf{A}_q \mathbf{r}[l]}_{\text{Multiuser interference}} + z_k[l],
\end{eqnarray}
where $z_{k}[l] \sim \mathcal{CN}(0,\sigma_{k}^2)$ denotes the additive white Gaussian noise (AWGN) at the $k$-th CU.
\par
\vspace*{2mm}
\noindent  \textit{A.2: Received Sensing Signal}
\vspace*{1mm}
\par
Assume the $q$-th AP serves as the sensing receiver. Then, the received echo reflected from the EHR can be given by
\begin{eqnarray} \label{sensing_signal}
\mathbf{y}_{q}[l] = \underset{q'\in\mathcal{Q}\setminus\{q\}}{\sum}\alpha_{q',q}\mathbf{G}_{q',q} \mathbf{x}_{q'}[l] + \mathbf{z}_{q}[l],
\end{eqnarray}
where vector $\mathbf{z}_q[l] \sim \mathcal{CN}(0,\sigma_{q}^2 \mathbf{I}_{N} )$ denotes the AWGN at the $q$-th AP. $\alpha_{q',q}$ is the radar cross section (RCS) of the link from the $q'$-th AP to the EHR and then to the $q$-th AP.
$\mathbf{G}_{q',q} \overset{\triangle}{=} \mathbf{a}(\theta_q)\mathbf{a}(\theta_{q'})^H \in \mathbb{C}^{N \times N}$ denotes the corresponding target response matrix, where $\theta_q$ ($\theta_{q'}$) is the AoA (AoD) of the EHR with respect to the $q$ ($q'$)-th AP. $\mathbf{a} (\theta_q) \in \mathbb{C}^N$ is the steering vector from the $q$-th AP to the EHR given by (assuming even number of antennas) \cite{9652071}
\begin{eqnarray}
\mathbf{a}(\theta_q) = [e^{-\frac{N-1}{2}\zeta_q}, e^{-\frac{N-3}{2}\zeta_q}, \cdots, e^{\frac{N-1}{2}\zeta_q}]^T,
\end{eqnarray}
where 
variable $\zeta_q$ is defined as $\zeta_q=j \pi \cos{\theta_q}$. Denote the 2-dimension Cartesian coordinates of the EHR and the $q$-th AP as $\mathbf{p}=[p_1, p_2]^T$ and $\mathbf{r}_q^{\mathrm{AP}} = [r_{q,1}^{\mathrm{AP}}, r_{q,2}^{\mathrm{AP}}]^T$, respectively, then $\cos{\theta_q}$ is given by
\begin{eqnarray}
    \cos{\theta_q} &\hspace*{-2mm}=\hspace*{-2mm}& \frac{r_{q,1}^{\mathrm{AP}}-p_{1}}{\sqrt{(r_{q,1}^{\mathrm{AP}}-p_{1})^2 + (r_{q,2}^{\mathrm{AP}}-p_{2})^2 }}.
\end{eqnarray}
\par
\subsection{Energy Harvesting Model}
To characterize the performance of the RF-based energy harvesting circuit,
we adopt the nonlinear saturation energy harvesting model described in \cite{7264986}
\begin{eqnarray}
& & P_{\mathrm{EH}} = \frac{\Psi-P_{\mathrm{sat}}\Xi}{1-\Omega}, \hspace{4mm} \Xi = \frac{1}{1+\exp{(ab)}}, \notag \\
& & \Psi=\frac{P_{\mathrm{sat}}}{1+\exp{(-a(P_{\mathrm{IN}}-b))}},
\end{eqnarray}
where $P_{\mathrm{EH}}$ denotes the average harvested power and $P_{\mathrm{sat}}$ represents the maximum harvested power when the energy harvesting circuit is saturated. $a$ and $b$ are constants related to the circuit characteristics.
Define $\mathbf{h}_{\mathrm{EH}} \overset{\triangle}{=} {\left[\mathbf{a}^T(\theta_1),\cdots, \mathbf{a}^T(\theta_Q) \right]}^T$ as the channel between APs and EHR for power transfer. Denote $\mathbf{R}$ as the covariance matrix of $\mathbf{r}[l]$. 
For a given AP selection $\mathbf{a}\overset{\triangle}{=}[a_1,\dots,a_Q]^T$, the received power at EHR can be expressed as
\begin{align} 
\label{input_power_10}
P_{\mathrm{IN}} = \underset{k\in\mathcal{K}}{\sum} \underset{q \in \mathcal{Q}}{\sum} (1-a_q) \left | \mathbf{h}_{\mathrm{EH}}^H 
\mathbf{A}_q \mathbf{w}_{k} \right |^2 \notag \\
+ \underset{q \in \mathcal{Q}}{\sum} (1-a_q) \mathbf{h}_{\mathrm{EH}}^H 
\mathbf{A}_q \mathbf{R} \mathbf{A}_q \mathbf{h}_{\mathrm{EH}}.
\end{align}
\par
Specifically, if the $q_0$-th AP is selected as the sensing receiver, i.e., $a_{q_0} = 0$ and $a_{q} = 1, \forall q \in \mathcal{Q} \setminus \{q_0\}$, then \eqref{input_power_10} can be expressed as
\begin{align}
P_{\mathrm{IN}} = \underset{k\in\mathcal{K}}{\sum} \left | \mathbf{h}_{\mathrm{EH}}^H 
\mathbf{A}_{q_0} \mathbf{w}_{k} \right |^2 + \mathbf{h}_{\mathrm{EH}}^H 
\mathbf{A}_{q_0} \mathbf{R} \mathbf{A}_{q_0} \mathbf{h}_{\mathrm{EH}}.
\end{align}
Hence, for any AP selection result, we can calculate the received power at EHR by \eqref{input_power_10}.
\section{Performance Metrics and Problem Formulation}
\subsection{Performance Metrics for Communication and Sensing}
\textbf{Communication Metric}: We utilize SINR to measure communication performance. According to \eqref{comm_signal_2}, 
the SINR for the $k$-th CU is given by
\begin{eqnarray}
\gamma_k(\mathbf{w}_k, \mathbf{R}) = \frac{P_{k,\mathrm{desired}}}{P_{k, \mathrm{user}} + P_{k,\mathrm{radar}} + \sigma_k^2},
\end{eqnarray}
where  
\begin{eqnarray}
P_{k,\mathrm{desired}} = \underset{q \in \mathcal{Q}}{\sum} (1-a_q) \left | \mathbf{h}_k^H \mathbf{A}_{q} \mathbf{w}_k \right |^2
\end{eqnarray}
denotes the power of the desired signal. Similarly, 
\begin{eqnarray}
P_{k,\mathrm{user}} = \underset{q \in \mathcal{Q}}{\sum} (1-a_q) \hspace{-3mm} \underset{k'\in\mathcal{K} \setminus \{k\}}{\sum} \hspace{-3mm} \left | \mathbf{h}_k^H 
\mathbf{A}_q \mathbf{w}_{k'}   \right |^2
\end{eqnarray}
represents the multi-user interference and 
\begin{eqnarray}
P_{k,\mathrm{radar}} = \underset{q \in \mathcal{Q}}{\sum} (1-a_q) \mathbf{h}_k^H \mathbf{A}_q \mathbf{R} \mathbf{A}_q \mathbf{h}_k
\end{eqnarray}
denotes the radar interference. 

\par
\vspace{1mm}
\textbf{Sensing Metric}: Following the common practice, we adopt CRB as the sensing performance metric, which is the lower bound for the mean square error (MSE) of any unbiased estimator \cite{9652071}.
In the following, we reorganize the received sensing signals to facilitate the derivation of the CRB.
First we rewrite \eqref{sensing_signal} as
\begin{eqnarray}
\mathbf{y}_{q}[l] = \mathbf{G}_q \mathbf{A}_q \widetilde{\mathbf{x}}[l] + \mathbf{z}_{q}[l],
\end{eqnarray}
where matrix $\mathbf{G}_q$ and vector $\widetilde{\mathbf{x}}[l]$ are defined as $\mathbf{G}_q = \left [ \alpha_{1,q} \mathbf{G}_{1,q},\cdots, \alpha_{Q,q} \mathbf{G}_{Q,q} \right ]$ and 
$\widetilde{\mathbf{x}}[l] = \left[ \widetilde{\mathbf{x}}_1[l]^T, \cdots, \widetilde{\mathbf{x}}_Q[l]^T \right]^T$, respectively, with
$\widetilde{\mathbf{x}}_q[l] \overset{\triangle}{=} \hspace{-1mm} \underset{k\in\mathcal{K}}{\sum} \mathbf{w}_{q,k}s_{k}[l] + \mathbf{r}_{q}[l] $. Note that $\widetilde{\mathbf{x}}_q[l]$ is obtained from $\mathbf{x}_q[l]$ by omitting the AP selection variable $a_q$.
The covariance matrix of $\widetilde{\mathbf{x}}[l]$ is given by
\begin{eqnarray} \label{covariance_S}
\mathbf{S} \overset{\triangle}{=} \mathbb{E}\left[ \widetilde{\mathbf{x}}[l] \widetilde{\mathbf{x}}[l]^H \right] = \underset{k \in \mathcal{K}}{\sum}
\mathbf{w}_k \mathbf{w}_k^H + \mathbf{R}.
\end{eqnarray}
\par
By stacking the $\eta L$ observed samples $\mathbf{y}_q[l]$, we have 
\begin{equation} \label{stack_Y}
\mathbf{Y}_{q} = \mathbf{G}_{q} \mathbf{A}_{q} \mathbf{X} + \mathbf{Z}_{q},
\end{equation}
where $\mathbf{Y}_{q} \overset{\triangle}{=} \left [ \mathbf{y}_{q}[1],\cdots, \mathbf{y}_{q}[\eta L]\right ]$, $\mathbf{X} \overset{\triangle}{=} \left [ \Tilde{\mathbf{x}}[1],\cdots, \Tilde{\mathbf{x}}[\eta L]\right ]$ and $\mathbf{Z}_{q} \overset{\triangle}{=} \left [ \mathbf{z}_{q}[1],\cdots, \mathbf{z}_{q}[\eta L]\right ]$.
The target parameter to be estimated is $\bm{\chi} = [\mathbf{p}^T, \bm{\alpha}_q^T]^T$, where
$\bm{\alpha}_q\in\mathbb{R}^{2(Q-1)\times 1}$ is given by 
\begin{eqnarray}
\bm{\alpha}_q\hspace*{-0.5mm}=\hspace*{-0.5mm}
\Big[\Re\{\alpha_{1,q}\},\cdots,\Re\{\alpha_{Q,q}\}, \Im\{\alpha_{1,q}\},\cdots,\Im\{\alpha_{Q,q}\}\Big]^T,
\end{eqnarray}
with $\Re\{\alpha_{q,q}\}, \Im\{\alpha_{q,q}\}$ excluded. The following proposition gives the CRB for location estimation in networked ISAC.
\newtheorem{prop}{Proposition}
\begin{prop}
Assume that the $q$-th AP is selected as the sensing receiver. Denote $\mathbf{F}_q \in \mathbb{R}^{2Q \times 2Q}$ as the Fisher information matrix (FIM) for estimating $\bm{\chi}$. Then the $(i,j)$-th element of $\mathbf{F}_q$ is given by
\begin{eqnarray} \label{prop_fim}
F_{ij}^{(q)} & = & \frac{2 \eta L}{\sigma_q^2} \Re \left\{ \mathrm{Tr} 
\left ( \dot{\mathbf{G}}^{(j)}_{q} \mathbf{A}_q \mathbf{S} \mathbf{A}_q \left [\dot{\mathbf{G}}^{(i)}_q  \right ] ^H  \right ) \right\}, \notag \\
& = & \frac{2 \eta L}{\sigma_q^2} \Re \left\{ \mathrm{vec} \left( \mathbf{S}\right)^H \bm{\lambda}_{ij}^{(q)} \right\},
\end{eqnarray}
where $\dot{\mathbf{G}}^{(i)}_{q}$ denotes the derivative of $\mathbf{G}_{q}$ with respect to $\bm{\chi}_i$ and $\bm{\lambda}_{ij}^{(q)} \overset{\triangle}{=} \mathrm{vec} \left( \mathbf{A}_q 
\left [\dot{\mathbf{G}}^{(j)}_{q} \right ] ^H \dot{\mathbf{G}}^{(i)}_q \mathbf{A}_q \right)$. The detailed derivation is given in Appendix A.
\par
Then, we partition $\mathbf{F}_q$ as
\begin{equation} \label{FIM_partition}
\mathbf{F}_q = \begin{bmatrix}
        \mathbf{F}^{(q)}_{\mathbf{p}\mathbf{p}} & \mathbf{F}^{(q)}_{\mathbf{p}\bm{\alpha}}\\
        \big[\mathbf{F}^{(q)}_{\mathbf{p}\bm{\alpha}}\big]^T & \mathbf{F}^{(q)}_{\bm{\alpha}\bm{\alpha}}\\
    \end{bmatrix},
\end{equation}
where $\mathbf{F}^{(q)}_{\mathbf{p}\mathbf{p}} \in \mathbb{R}^{2 \times 2}$. Then,
the corresponding CRB matrix for estimating the EHR's location $\mathbf{p}$ based on the observation $\mathbf{Y}_q$ is given by \cite{1703855}
\begin{equation} \label{CRB}
\mathbf{CRB}^{(q)}(\mathbf{p}) = \left[ \mathbf{F}_{\mathbf{p}\mathbf{p}}^{(q)} - \mathbf{F}^{(q)}_{\mathbf{p}\bm{\alpha}} \left[\mathbf{F}^{(q)}_{\bm{\alpha}\bm{\alpha}}\right]^{-1} \left[ \mathbf{F}^{(q)}_{\mathbf{p}\bm{\alpha}} \right]^T \right] ^{-1}.\\
\end{equation}
\end{prop}
The CRB for estimating $p_i$ is given by 
\begin{eqnarray}
\mathrm{CRB}^{(q)}(p_i) = \left[ \mathbf{CRB}^{(q)}(\mathbf{p}) \right]_{i,i}.
\end{eqnarray}
The diagonal element of the CRB matrix denotes the variance of the corresponding parameter to be estimated.
\par
Due to the movement of the EHR and the sensing estimation error, we may not be able to acquire the accurate location of the EHR, leading to unavoidable location uncertainty. To capture this effect, we adopt a widely used uncertainty model with a deterministic bound \cite{9933849}, given by 
\begin{eqnarray}
\mathbf{p} = \overline{\mathbf{p}} + \Delta \mathbf{p},
\end{eqnarray}
where $\mathbf{p}$ and $\overline{\mathbf{p}} \overset{\triangle}{=} [\overline{p}_1, \overline{p}_2]^T$ are the true location and estimated location of the EHR, respectively. Moreover, $\Delta \mathbf{p} \overset{\triangle}{=} [\Delta p_1, \Delta p_2]^T$ represents the location error and the two components of $\Delta \mathbf{p}$, i.e., $\Delta p_1$ and $\Delta p_2$, are assumed to be upper bounded with $\left | \Delta p_1 \right | 
\leq \psi_1$ and $\left | \Delta p_2 \right | \leq \psi_2$, respectively. We further define the location uncertainty area $\Omega$ as
\begin{eqnarray} \label{uncertainty_error}
\Omega \overset{\triangle}{=} \left\{ (\Delta p_1, \Delta p_2) \mid \left| \Delta p_i \right| \leq \psi_i , i \in \left\{ 1,2\right\} \right\}.
\end{eqnarray}
\subsection{Problem Formulation}
At the beginning of each time frame, the channels between users and APs are acquired at each AP and reported to the central controller. The EHR and CUs report their energy harvesting requirement and SINR requirement to the central controller, respectively. Based on the above information, the central controller designs the beamforming policy and the AP selection strategy. A two-phase design is proposed to minimize power consumption while localizing and transferring power to the EHR, as well as serving the CUs. \footnote{A multi-objective optimization problem (MOOP) can be formulated to investigate the trade-offs between sensing performance and power consumption. In this paper, our main objective is to investigate the use of sensing to assist SWIPT, where we are particularly interested in the energy efficient design. As a result, we formulate the problem as a two-phase power minimization problem instead of a multi-objective design.} This is achieved by optimizing the time-splitting ratio, beamforming vectors, radar covariance matrix, and AP selection. \footnote{
The final AP selection result can be indicated by the solution of $a_q$ after solving the optimization problem. Specifically, the $q$-th AP is assigned as the transmitter if $a_q=1$. Otherwise, the $q$-th AP is assigned as the sensing receiver.} Note that the two phases involve two different optimization problems, both of which are related to the time-splitting ratio. In addition, the WPT phase relies on the results of the sensing phase. 
To this end, we propose a two-layer optimization framework. In particular, in the inner layer, we optimize the sensing and WPT phases with a given time-splitting ratio $\eta$. Then, in the outer layer, we use line search to determine $\eta$. In the following, we first present the two sub-problems associated with the sensing phase and the WPT phase.
\par
\textbf{Sensing phase}: The sensing phase aims to minimize the CRB for estimating the location of the EHR in the presence of location uncertainty. For that purpose, we jointly optimize the beamforming policy and the AP selection. The corresponding optimization problem is formulated as follows
\begin{eqnarray} \label{step1}
&&\hspace{-10mm}\underset{\substack{ \mathbf{w}_k^{\text{I}}, \mathbf{R}^{\text{I}}, a_q }}{\mino} \hspace*{4mm}\underset{\Delta \mathbf{p} \in \Omega}{\max} \hspace*{1mm} \mathrm{Tr} \left ( \mathbf{CRB} (\mathbf{p}) \right) \notag\\
&&\hspace*{-10mm}\subto\hspace*{1mm} \mbox{C1:}\hspace*{1mm} a_q \Big (\hspace*{-0.5mm}\underset{k \in \mathcal{K}}{\sum} \left\Vert \mathbf{B}_q \mathbf{w}_k^{\text{I}} \right\Vert_2^2\hspace*{-0.5mm}+ \hspace*{-0.5mm}\mathrm{Tr} \big(\mathbf{B}_q \mathbf{R}^{\text{I}} \big)\hspace*{-0.5mm}\Big )\hspace*{-0.5mm}\leq \hspace*{-0.5mm}P_{\mathrm{max}},\hspace*{1mm} \forall q, \notag \\
&&\hspace*{8mm} \mbox{C2:}\hspace*{1mm} \gamma_{k}(\mathbf{w}_k^{\text{I}}, \mathbf{R}^{\text{I}}) \geq \Gamma_{k}, \hspace*{1mm} \forall k, \notag \\
&&\hspace*{8mm} \mbox{C3:}\hspace*{1mm} \underset{q \in \mathcal{Q}}{\sum} a_q = Q-1, \notag \\
&&\hspace*{8mm} \mbox{C4:}\hspace*{1mm} a_q \in \left \{ 0, 1 \right \}, \hspace*{1mm} \forall q,
\end{eqnarray}
where
\begin{align} \label{all_crb}
\mathbf{CRB} (\mathbf{p}) = \underset{q \in \mathcal{Q}}{\sum} (1-a_q) \mathbf{CRB}^{(q)}(\mathbf{p}).
\end{align}
For any AP $q \in \mathcal{Q}$ selected as the sensing receiver, $\mathbf{CRB} (\mathbf{p}) = \mathbf{CRB}^{(q)}(\mathbf{p})$ denotes the CRB matrix for estimating the EHR's location. Hence, for any given AP selection result, the sensing performance can be evaluated by \eqref{all_crb}.
Here, the optimization variables $\mathbf{w}_k^{\text{I}}$ and $\mathbf{R}^{\text{I}}$ denote the beamforming vector for CU $k$ and radar covariance matrix in the sensing phase, respectively. $P_{\mathrm{max}}$ in constraint C1 indicates the maximum available transmit power of each AP. Constraint C2 ensures that the SINR of the $k$-th CU is larger than a required SINR threshold $\Gamma_k$. Constraints C3 and C4 are imposed to guarantee that one AP is selected as the sensing receiver. Next, we further explain constraint C1. Assume that the $q_1$-th AP is selected as the sensing receiver, i.e., $a_{q_1}=0$. Then for any $q \neq q_1$, we have $a_q = 1$ and
$\underset{k \in \mathcal{K}}{\sum} \left\Vert \mathbf{B}_q \mathbf{w}_k^{\text{I}} \right\Vert_2^2+ \mathrm{Tr} \big(\mathbf{B}_q \mathbf{R}^{\text{I}} \big) \leq P_{\mathrm{max}}$ where $\mathbf{B}_q \overset{\triangle}{=} \mathbf{I} - \mathbf{A}$. This limits the maximum transmit power of each transmitter. If $q=q_1$ and $a_{q}=0$, then constraint C1 is satisfied for any non-negative $P_{\mathrm{max}}$.
\par
\textbf{WPT phase}: The refined location information of the EHR and the AP selection strategy obtained in the sensing phase are exploited to facilitate efficient SWIPT during the WPT phase. In particular, we aim to minimize the average system transmit power while guaranteeing the SINR requirement of the CUs and the minimum harvested energy requirement of the EHR. We solve for the energy-efficient beamforming policy by the following optimization problem
\begin{eqnarray} \label{step2}
&&\hspace{-10mm}
\underset{\substack{ \mathbf{w}_k^{\text{II}}, \mathbf{R}^{\text{II}}} }{\mino} \hspace*{2mm} P_{\mathrm{avg}} \notag\\
&&\hspace*{-10mm}\subto \hspace*{1mm} \mbox{C5:}\hspace*{1mm} a_q \Big (\hspace*{-0.5mm}\underset{k \in \mathcal{K}}{\sum} \left\Vert \mathbf{B}_q \mathbf{w}_k^{\text{II}} \right\Vert_2^2\hspace*{-0.5mm}+\hspace*{-0.5mm}\mathrm{Tr} \big(\mathbf{B}_q \mathbf{R}^{\text{II}} \big) \Big )\hspace*{-0.5mm}\leq\hspace*{-0.5mm}P_{\mathrm{max}}, \hspace*{1mm}\forall q, \notag \\
&&\hspace*{8mm} \mbox{C6:}\hspace*{1mm} \gamma_{k}(\mathbf{w}_k^{\text{II}}, \mathbf{R}^{\text{II}}) \geq \Gamma_{k}, \hspace*{1mm} \forall k, \notag \\
&&\hspace*{8mm} \mbox{C7:} \hspace*{1mm} \eta P_{\mathrm{EH}}^{\mathrm{I}} + (1-\eta) \hspace*{1mm} \underset{\Delta \mathbf{p}^+ \in \Omega^+ }{\min} P_{\mathrm{EH}}^{\mathrm{II}} \geq P_{\mathrm{req}},
\end{eqnarray}
where $\mathbf{w}_k^{\text{II}}$ and $\mathbf{R}^{\text{II}}$ denote the beamforming vector for CU $k$ and radar covariance matrix in the sensing phase, respectively. The objective function $P_{\mathrm{avg}}$ is the average power of the whole frame given by
\begin{eqnarray}
P_{\mathrm{avg}} &\hspace*{-2mm}\overset{\triangle}{=}\hspace*{-2mm}&  \eta \underset{q \in \mathcal{Q}}{\sum} a_q \Big ( \underset{k \in \mathcal{K}}{\sum} \left \| \mathbf{B}_q \mathbf{w}_k^{\text{I}} \right \|^2_2 + \mathrm{Tr}\big( \mathbf{B}_q \mathbf{R}^{\text{I}} \big) \Big ) \notag \\
&\hspace*{-2mm}+\hspace*{-2mm}& \left ( 1-\eta \right ) \underset{q \in \mathcal{Q}}{\sum} a_q \Big( \underset{k \in \mathcal{K}}{\sum} \left \| \mathbf{B}_q \mathbf{w}_k^{\text{II}} \right \|^2_2 + \mathrm{Tr}\big( \mathbf{B}_q \mathbf{R}^{\text{II}} \big) \Big).
\end{eqnarray}
Moreover, constraint C5 limits the power consumption of each AP to be less than the maximum available power $P_{\mathrm{max}}$. Constraint C6 ensures that the SINR of the $k$-th CU is larger than the minimum SINR requirement $\Gamma_k$. Constraint C7 is imposed to satisfy the harvested energy requirement, i.e., $P_{\text{req}}$. Specifically, 
$P_{\mathrm{EH}}^{\mathrm{I}}$ and $P_{\mathrm{EH}}^{\mathrm{II}}$ in C7 denote the harvested energy of the EHR during the sensing phase and WPT phase, respectively.
Since the actual value of $P_{\mathrm{EH}}^{\mathrm{I}}$ cannot be obtained due to the unknown true location, we use the safe approximation of $P_{\mathrm{EH}}^{\mathrm{I}}$. In particular, given $\mathbf{w}_k^{\mathrm{I}}$, $\mathbf{R}^{\mathrm{I}}$ and $a_q$ obtained in the sensing phase, we evaluate the minimum possible value of $P_{\mathrm{EH}}^{\mathrm{I}}$ within the potential uncertainty area, i.e.,
$\underset{\substack{ \Delta \mathbf{p} \in \Omega }}{\mino} \hspace{2mm} P_{\mathrm{EH}}^{\mathrm{I}}$. This can be solved by quantizing the position in the location uncertainty area $\Omega$ and choosing the minimum value over the quantized positions. Given the location uncertainty is normally a small value, this can be solved with low complexity. 
$\Delta \mathbf{p}^+ \overset{\triangle}{=} [\Delta p^+_1, \Delta p^+_2]^T$ in C7 denotes the refined location estimation error obtained from the sensing phase. According to the three-sigma rule of thumb, $\Delta p^+_i$ is bounded by a three-fold the square root of $\mathrm{CRB}(p_i)$, i.e.,
$\left| \Delta p_i^+ \right| \leq \psi_i^{+} \overset{\triangle}{=} 3\sqrt{\mathrm{CRB}(p_i)}$ \cite{10227884, 10375133}.
$\Omega^{+}$ is the refined location uncertainty area defined in \eqref{uncertainty_error} with the updated location error.
\par
Note that both problem \eqref{step1} and problem \eqref{step2} are non-convex and difficult to solve. For problem \eqref{step1}, the non-convexity originates from the complicated form of the CRB and the location uncertainty contained in the CRB matrix. The AP selection variables $a_q$ have a tricky binary form and are coupled with the beamforming policy $\mathbf{w}_k^{\text{I}}$ and $\mathbf{R}^{\text{I}}$. For problem \eqref{step2}, the non-convexity comes from the complicated form of the energy harvesting model and the corresponding robust design in constraint C7. Note that the time-splitting ratio is involved in both sub-problems. In addition, the problem \eqref{step2} relies on the optimization results of problem $\eqref{step1}$. These make the problem very challenging to solve. 
\begin{figure*}[t]
\centering
\includegraphics[width=6.3in]{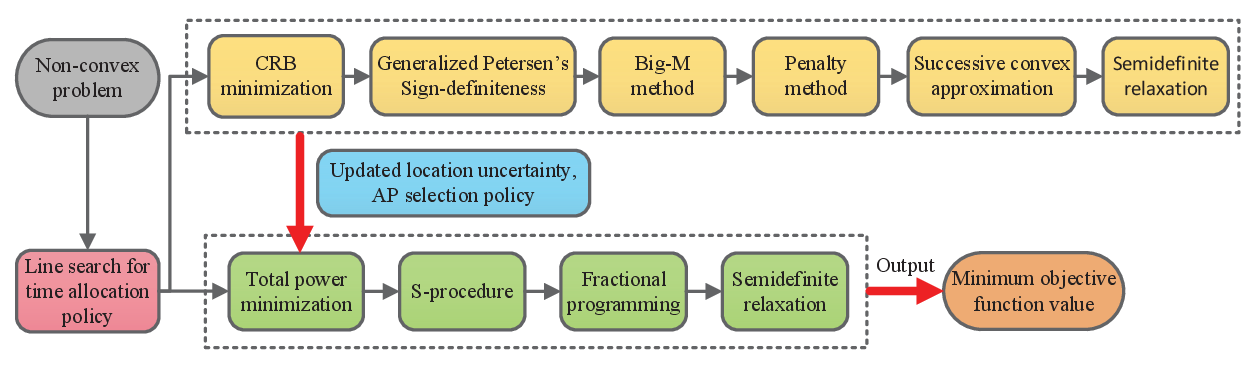}\caption{Illustration of the key steps of the proposed two-layer algorithm.}
\label{fig: flowchart}
\end{figure*}
\section{Solution of the Optimization Problem}
In this section, we develop a two-layer optimization framework. In particular, in the outer layer, we use the line search method to traverse all possible time-splitting strategies. In the inner layer, we sequentially solve the two sub-problems for each time allocation strategy. Finally, the solution that produces the minimum objective function value of the problem \eqref{step2} is selected. The proposed algorithm solves the problem with polynomial computational complexity and guaranteed convergence. The key steps of the proposed optimization framework are presented in Fig. \ref{fig: flowchart}.
\par
Note that the optimization of $\mathbf{w}_k^{\text{I}}$ and $\mathbf{w}_k^{\text{II}}$ are independent and separated in problem \eqref{step1} and problem \eqref{step2}, respectively. For notational simplicity, we slightly abuse the notation $\mathbf{w}_k$ to denote $\mathbf{w}_k^{\text{I}}$ and $\mathbf{w}_k^{\text{II}}$ when solving problem \eqref{step1} and problem \eqref{step2}, respectively. Similarly, $\mathbf{R}$ is used to represent respectively $\mathbf{R}^{\text{I}}$ and $\mathbf{R}^{\text{II}}$ when solving two problems.
\subsection{Solution of Problem \eqref{step1}}
To facilitate the problem solving, we first transform the CRB matrix into a more tractable form. Then, the location uncertainty in each term of the FIM is circumvented by employing the Generalized Petersen’s Sign-definiteness Lemma. Subsequently, the big-M method is applied to decouple the binary variable $a_k$ and the beamforming variables $\mathbf{w}_k^{\text{I}}$ and $\mathbf{R}^{\text{I}}$. Finally, 
by sequentially capitalizing on the penalty method, SCA, and SDR, we overcome the non-convexity of problem \eqref{step1}.
\par
To start with, we define matrices $\mathbf{H}_k \overset{\triangle}{=} \mathbf{h}_k \mathbf{h}_k^H$ and $\mathbf{W}_k \overset{\triangle}{=} \mathbf{w}_k \mathbf{w}_k^H$, $\forall k$. Then, the covariance matrix $\mathbf{S}$ in \eqref{covariance_S} can be equivalently expressed as $\mathbf{S} = \underset{k \in \mathcal{K}}{\sum}
\mathbf{W}_k + \mathbf{R}$.
Also, constraints C1 and C2 can be rewritten equivalently as 
\begin{eqnarray}
\overline{\mbox{C1}}\mbox{:} \hspace{2mm} a_q \left ( \underset{k \in \mathcal{K}}{\sum} \mathrm{Tr} \left( \mathbf{B}_q \mathbf{W}_k \right) + \mathrm{Tr} \left \{ \mathbf{B}_q \mathbf{R} \right \} \right ) \leq P_{\mathrm{max}}, \hspace*{1mm} \forall q
\end{eqnarray}
and
\begin{eqnarray}
\hspace{-5mm} \overline{\mbox{C2}} \mbox{:} && \hspace{-5mm} \frac{1}{\Gamma_k}  \underset{q \in \mathcal{Q}}{\sum} (1-a_q)  \mathrm{Tr} \left (  \mathbf{A}_q \mathbf{H}_k \mathbf{A}_q \mathbf{W}_k \right ) \notag \\
&& \hspace{-5mm} \geq \underset{q \in \mathcal{Q}}{\sum} (1-a_q) \hspace{-3mm} \underset{k'\in\mathcal{K} \setminus \{k\}}{\sum} \hspace{-3mm} \mathrm{Tr} \left ( \mathbf{A}_q \mathbf{H}_k 
\mathbf{A}_q \mathbf{W}_{k'} \right ) \notag \\
&& \hspace{-5mm} + \underset{q \in \mathcal{Q}}{\sum} (1-a_q) \mathrm{Tr} \left ( \mathbf{A}_q \mathbf{H}_k 
\mathbf{A}_q \mathbf{R} \right ) + \sigma_k^2, \hspace{1mm} \forall k,
\end{eqnarray}
respectively.
\par
As a result, the problem \eqref{step1} can be recast equivalently as follows
\begin{eqnarray} \label{step1_2}
&&\hspace{-17mm}\underset{\substack{ \mathbf{W}_k, \mathbf{R}, a_q }}{\mino} \hspace*{5mm}\underset{\Delta \mathbf{p}}{\max} \hspace*{1mm} \mathrm{Tr} \left ( \mathbf{CRB} (\mathbf{p}) \right) \notag\\
&&\hspace*{-17mm}\subto\hspace*{2mm} \overline{\mbox{C1}}, \overline{\mbox{C2}}, \mbox{C3}, \mbox{C4}, \mbox{C8:}\hspace*{1mm} \mathrm{Rank}\left( \mathbf{W}_k \right) = 1,
\end{eqnarray}
where the rank-one constraint C8 is imposed to guarantee the equivalent recovery of $\mathbf{w}_k$ for problem \eqref{step1} from the $\mathbf{W}_k$ obtained in \eqref{step1_2}.
\par
Next, we aim to transform the CRB in the objective function into a more tractable form. Note that the function $\mathrm{Tr}\left(\mathbf{X}^{-1}\right)$ is decreasing on the positive semidefinite matrix space. Based on this observation, we introduce an auxiliary optimization variable $\mathbf{J} \in \mathbb{C}^{2 \times 2} \succeq \mathbf{0}$ and rewrite the problem in \eqref{step1_2}
equivalently as the problem \eqref{crb_trans} which is shown at the top of the next page.
\begin{remark}
The proposed method can be extended to the case with multiple
mobile EHRs but needs some modifications in the sensing phase. Specifically, we assume the system serves $M$ EHRs and define set $\mathcal{M}\overset{\Delta }{=}\left\{1,\cdots,M\right\}$ to collect the indices of the EHRs. Under such circumstances, the sensing phase aims to minimize the maximum CRB among all EHRs, which involves a mini-max problem. To solve this, we can introduce auxiliary variables $\mathbf{J}_m, m\in \mathcal{M}$ and $t \geq 0$. Then, the problem can be reformulated by replacing the objective function in \eqref{crb_trans} with $t$ and introducing additional constraint $\mathrm{Tr}(\mathbf{J}_m) \leq t, \hspace*{1mm} \forall m$.
\end{remark}
\begin{figure*}[t]
\setcounter{TempEqCnt}{\value{equation}} 
\setcounter{equation}{30} 
\begin{eqnarray} \label{crb_trans}
&&\hspace{-7mm}\underset{\substack{ \mathbf{W}_k, \mathbf{R}, a_q }}{\mino} \hspace*{4mm} \mathrm{Tr} \left ( \mathbf{J}^{-1} \right) \notag\\
&&\hspace*{-8mm}\subto\hspace*{2mm} \overline{\mbox{C1}}, \overline{\mbox{C2}}, \mbox{C3}, \mbox{C4}, \mbox{C8}, \mbox{C9:}\hspace*{1mm}\mathbf{J} \succeq \mathbf{0}, \notag \\
&&\hspace*{11mm} \mbox{C10:} \hspace*{1mm} \hspace*{1mm} (1-a_q)\left( \mathbf{F}_{\mathbf{p}\mathbf{p}}^{(q)} - \mathbf{F}^{(q)}_{\mathbf{p}\bm{\alpha}} \left[\mathbf{F}^{(q)}_{\bm{\alpha}\bm{\alpha}}\right]^{-1} \mathbf{F}^{(q)}_{\mathbf{p}\bm{\alpha}}-\mathbf{J} \right)\succeq \mathbf{0}, \hspace*{1mm}\forall q, \hspace*{1mm}\Delta \mathbf{p} \in \Omega.
\end{eqnarray}
\hrulefill
\vspace{-6mm}
\end{figure*}
\setcounter{equation}{31}
\par
We note that constraint C10 is still non-convex due to the product of the submatrices and the uncertainty in the location of the EHR. To circumvent this difficulty, we first use the Schur complement \cite{zhang2006schur} to equivalently transform C10 into
\vspace{1mm}
\begin{eqnarray}
\overline{\mbox{C10}} \mbox{:} \hspace*{1mm}
(1-a_q)
\begin{bmatrix}
\mathbf{F}_{\mathbf{p}\mathbf{p}}^{(q)} - \mathbf{J} & \mathbf{F}^{(q)}_{\mathbf{p}\bm{\alpha}} \\
\left[ \mathbf{F}^{(q)}_{\mathbf{p}\bm{\alpha}} \right]^T & \mathbf{F}^{(q)}_{\bm{\alpha}\bm{\alpha}}
\end{bmatrix} \succeq \mathbf{0}, \hspace*{1mm}\forall q, \hspace*{1mm}\Delta \mathbf{p} \in \Omega.
\end{eqnarray}
\vspace{1mm}
Then, we focus on dealing with the uncertainty laid in $\overline{\mbox{C10}}$. As shown in \eqref{prop_fim}, the uncertainty originates from the term $\bm{\lambda}_{ij}^{(q)}$ which includes the geometry information of APs and the EHR. However, $\bm{\lambda}_{ij}^{(q)}$ is nonlinear with respect to $\Delta \mathbf{p}$ which makes the resource allocation design very challenging. To circumvent this difficulty, we use the first-order Taylor expansion to approximate $\bm{\lambda}_{ij}^{(q)}$ \cite{8974403}
\begin{eqnarray} \label{taylor_approx}
\bm{\lambda}_{ij}^{(q)} = \overline{\bm{\lambda}}_{ij}^{(q)} + \dot{\bm{\lambda}}^{(q)}_{ij,1} \Delta p_1 + \dot{\bm{\lambda}}^{(q)}_{ij,2} \Delta p_2,
\end{eqnarray}
where variable $\overline{\bm{\lambda}}_{ij}^{(q)}$ is the estimated value of $\bm{\lambda}_{ij}^{(q)}$ at $\overline{\mathbf{p}}$. Also,
variables $\dot{\bm{\lambda}}^{(q)}_{ij,1}$ and $\dot{\bm{\lambda}}^{(q)}_{ij,2}$ denote the derivatives of $\bm{\lambda}^{(q)}_{ij}$ with respect to $p_1$ and $p_2$, respectively. 
Define $ \Delta \bm{\lambda}_{ij}^{(q)} = \left [ \left [ \dot{\bm{\lambda}}^{(q)}_{ij,1} \right]^T   \Delta p_1, \left [ \dot{\bm{\lambda}}^{(q)}_{ij,2} \right ]^T \Delta p_2 \right ]^T $. Then $\Delta \bm{\lambda}_{ij}^{(q)}$ is upper bounded by
\begin{eqnarray}
\left \| \Delta \bm{\lambda}^{(q)}_{ij} \right \|
&\hspace{-2mm}= \hspace{-2mm}& \left( \left \| \dot{\bm{\lambda}}^{(q)}_{ij,1} \right \|^2 \Delta p_1^2 + \left \| \dot{\bm{\lambda}}^{(q)}_{ij,2} \right \|^2 \Delta p_2^2 \right)^{\frac{1}{2}} \notag \\
&\hspace{-2mm}\leq\hspace{-2mm}& \left( \left \| \dot{\bm{\lambda}}^{(q)}_{ij,1} \right \|^2 \psi_1^2 + \left \| \dot{\bm{\lambda}}^{(q)}_{ij,2} \right \|^2 \psi_2^2 \right)^{\frac{1}{2}}  \overset{\Delta}{=} \varrho_{ij}^{(q)},
\end{eqnarray}
where $\varrho_{ij}^{(q)}$ is defined as the error bound of $\Delta \bm{\lambda}_{ij}^{(q)}$ caused by the location uncertainty.
Define $\Tilde{\mathbf{s}} \overset{\Delta}{=} \mathrm{vec} \left(\mathbf{S} \right) \in \mathbb{R}^{N^2Q^2}$, $\overline{\mathbf{s}} \overset{\Delta}{=} \left [ 1,1 \right]^T \otimes  \Tilde{\mathbf{s}}$, $\myoverline{\mathbf{s}} \overset{\Delta}{=} \left[
\Re\left\{\overline{\mathbf{s}}\right\}^T, \Im\left\{\overline{\mathbf{s}}\right\}^T
\right]^T$, and $\Delta \widetilde{\bm{\lambda}}^{(q)}_{ij} = \left[ \Re\left\{\Delta \bm{\lambda}^{(q)}_{ij}\right\}^T, \Im\left\{\Delta \bm{\lambda}^{(q)}_{ij}\right\}^T \right]^T \in \mathbb{R}^{4N^2Q^2}$. Note that $\Delta \widetilde{\bm{\lambda}}^{(q)}_{ij}$ has the same error bound as $\Delta \bm{\lambda}^{(q)}_{ij}$. Then, we rewrite the elements of the FIM in \eqref{prop_fim} as follows
\vspace{1mm}
\begin{eqnarray} \label{wl}
F_{ij}^{(q)} && \hspace{-6mm} = \frac{2 \eta L}{\sigma_q^2} \Re \{\Tilde{\mathbf{s}}^H \bm{\lambda}^{(q)}_{ij}\} = \hspace{-1mm} \frac{2 \eta L}{\sigma_q^2} \hspace{-1mm} \left( \Re \{\Tilde{\mathbf{s}}^H \bar{\bm{\lambda}}^{(q)}_{ij}\} + \Re \{\overline{\mathbf{s}}^H \Delta \bm{\lambda}^{(q)}_{ij} \} \right) \notag \\
&& \hspace{-6mm} = \frac{2 \eta L}{\sigma_q^2} \left( \Re \{\Tilde{\mathbf{s}}^H \bar{\bm{\lambda}}^{(q)}_{ij}\} + \myoverline{\mathbf{s}}^H \Delta \widetilde{\bm{\lambda}}^{(q)}_{ij} \right).
\end{eqnarray}
By substituting \eqref{wl} into constraint $\overline{\mbox{C10}}$, we have
\begin{eqnarray} \label{_C10}
\overline{\mbox{C10}} \mbox{:}
(1-a_q) \left( 
\overline{\mathbf{F}}^{(q)} 
-
\begin{bmatrix}
\mathbf{J} & \mathbf{0} \\
\mathbf{0} & \mathbf{0}
\end{bmatrix}
\right)
\succeq 
-(1-a_q)
\Delta \mathbf{F}^{(q)}, \notag \\
\forall q, \hspace{1mm} \Delta \mathbf{p} \in \Omega,
\end{eqnarray}
where the $(i,j)$-th elements of $\overline{\mathbf{F}}^{(q)}$ and $\Delta \mathbf{F}^{(q)}$ are given by $ \frac{2 \eta L}{\sigma_q^2} \Re \{\Tilde{\mathbf{s}}^H \bar{\bm{\lambda}}^{(q)}_{ij}\}$ and $\frac{2 \eta L}{\sigma_q^2} \myoverline{\mathbf{s}}^H \Delta \widetilde{\bm{\lambda}}^{(q)}_{ij}$, respectively.
In the following, we tackle the uncertainty contained in the elements of FIM by exploiting the following lemma.
\begin{lemma}
\textit{(Generalized Petersen’s Sign-definiteness Lemma \cite{6320700})}
Given matrices $\mathbf{M}$ and $\left\{ \mathbf{P}_i, \mathbf{Q}_i \right\}_{i=1}^{M}$, the semi-infinite linear matrix inequality (LMI) of the form
\begin{eqnarray} \label{lemma1_1}
\mathbf{M} \succeq \sum_{i=1}^{M}  \left (  \mathbf{P}_i^H\mathbf{X}_i\mathbf{Q}_i + \mathbf{Q}_i^H\mathbf{X}_i^H\mathbf{P}_i \right ), \hspace{1mm} \forall i, \mathbf{X}_i: \left \| \mathbf{X}_i \right \| \leq \aleph_i,
\end{eqnarray}
holds if and only if there exist nonnegative real numbers $\epsilon_1,\dots,\epsilon_M$ such that
\begin{eqnarray}
\begin{bmatrix}
\mathbf{M}- \sum_{i=1}^{M} \epsilon_i \mathbf{Q}_i^H \mathbf{Q}_i  & -\aleph_i \mathbf{P}_1^H & \cdots & -\aleph_N \mathbf{P}_N^H \\
-\aleph_i \mathbf{P}_1 & \epsilon_1 \mathbf{I} & \cdots & \mathbf{0} \\
\vdots & \vdots & \ddots & \vdots \\
-\aleph_N \mathbf{P}_N & \mathbf{0} & \cdots & \epsilon_M \mathbf{I}
\end{bmatrix}
\succeq \mathbf{0}.
\end{eqnarray}
The matrix norm in \eqref{lemma1_1} is the Frobenius norm.
\end{lemma}
\par
To facilitate the application of $\textbf{Lemma 1}$,
the RHS of \eqref{_C10} is reformulated as
\begin{eqnarray}
\mathrm{RHS}_{\overline{\mathrm{C10}}} = \hspace{-6mm} && (1-a_q) 
\underset{i,j}{\sum} 
\Big(
\mathbf{P}_{ij}^H \Delta \widetilde{\bm{\lambda}}_{ij}^{(q)} \mathbf{q}_{ij} + \mathbf{q}_{ij}^H  [\Delta \widetilde{\bm{\lambda}}^{(q)}_{ij}]^H \mathbf{P}_{ij}
\Big), \notag \\
&& \hspace{18mm} (i,j) \in \left\{ (i,j) \mid i,j \in \mathcal{Q}, j \geq i \right\},
\end{eqnarray}
where $\mathbf{P}_{ij} \in \mathbb{R}^{4N^2Q^2 \times 2Q}$. Note that the $i$-th column of $\mathbf{P}_{ij}$ is $\myoverline{\mathbf{s}}$ while the other components are $0$. Moreover, $\mathbf{q}_{ij} \in \mathbb{R}^{1 \times 2Q}$ is a vector and  the $i$-th element of $\mathbf{q}_{ij}$ is $-\frac{1}{2}$ if $i=j$. If $i \neq j$, then $j$-th element of $\mathbf{q}_{ij}$ is $-1$. The other terms of $\mathbf{q}_{ij}$ are $0$.
\par
Then, we introduce auxiliary optimization variables $\epsilon_{ij}$ where $\epsilon_{ij} \geq 0, (i,j) \in \left\{ (i,j) \mid i,j \in \mathcal{Q}, j \geq i \right\}$, and apply $\textbf{Lemma 1}$ to equivalently recast constraint $\overline{\mbox{C10}}$ as follows
\begin{eqnarray}
\myoverline{\mbox{C10}} \mbox{:}
(1-a_q)\begin{bmatrix}
\mathbf{M}- \mathbf{Q}  & \mathbf{P}^H \\
\mathbf{P} & \mathbf{T}
\end{bmatrix}
\succeq \mathbf{0}, \forall q,
\end{eqnarray}
where $\mathbf{M} = 
\overline{\mathbf{F}}^{(q)} 
-
\begin{bmatrix}
\mathbf{J} & \mathbf{0} \\
\mathbf{0} & \mathbf{0}
\end{bmatrix}.$
Matrices $\mathbf{Q}$, $\mathbf{P}$, and $\mathbf{T}$ are given by, respectively,
\begin{eqnarray}
\mathbf{Q} \hspace{-2mm}& = \hspace{-2mm}& \underset{i,j}{\sum}
\epsilon_{ij} \mathbf{q}_{ij}^H\mathbf{q}_{ij}, \\
\mathbf{P} \hspace{-2mm}& = \hspace{-2mm}& \left[ -\varrho^{(q)}_{11} \mathbf{P}_{11}^T, \cdots, -\varrho^{(q)}_{ij} \mathbf{P}_{ij}^T, \cdots, -\varrho^{(q)}_{QQ} \mathbf{P}_{QQ}^T \right]^T \hspace{-1mm}, \\
\mathbf{T} \hspace{-2mm}& = \hspace{-2mm}& \mathrm{\diag} \left\{ \left[ \epsilon_{11}, \cdots, \epsilon_{ij}, \cdots, \epsilon_{QQ} \right] \right\} \otimes \mathbf{I}_{4N^2Q^2}.
\end{eqnarray}
\par
Note that the AP selection variables $a_q$ are coupled with the beamforming policy $\mathbf{W}_k$, $\mathbf{R}$, and auxiliary variables $\mathbf{J}$, $\epsilon_{ij}^{(q)}$ in constraints $\overline{\mbox{C1}}$, $\overline{\mbox{C2}}$, and $\myoverline{\mbox{C10}}$. To deal with this issue, we propose to use the big-M method \cite{griva2009linear} to decouple the optimization variables by defining new optimization variables $\overline{\mathbf{W}}_{q,k} = a_q \mathbf{W}_{k}, \forall q,k$, $\overline{\mathbf{R}}_q = a_q \mathbf{R}, \forall q$, $\overline{\mathbf{J}}_{q} = a_q \mathbf{J}, \forall q$, resulting in the following constraints
\begin{eqnarray}
&& \hspace*{-12mm} \mbox{C11a:}\hspace*{1mm}\overline{\mathbf{W}}_{q,k} \preceq a_{q} (Q-1) P_{\mathrm{max}} \mathbf{I},\hspace*{1mm}\forall q,\hspace*{1mm}\forall k,\\
&& \hspace*{-12mm} \mbox{C11b:}\hspace*{1mm}\overline{\mathbf{W}}_{q,k} \succeq \mathbf{W}_{k}-(1-a_{q})(Q-1) P_{\mathrm{max}} \mathbf{I},\hspace*{1mm}\forall q,\hspace*{1mm}\forall k,\\
&& \hspace*{-12mm} \mbox{C11c:}\hspace*{1mm}\overline{\mathbf{W}}_{q,k} \preceq \mathbf{W}_{k},\hspace*{1mm}\forall q,\hspace*{1mm}\forall k,\hspace*{1mm} \mbox{C11d:}\hspace*{1mm}\overline{\mathbf{W}}_{q,k} \succeq \mathbf{0},\hspace*{1mm}\forall q,\hspace*{1mm}\forall k.
\end{eqnarray}
These four constraints are equivalent to characterize the relationship of $\overline{\mathbf{W}}_{q,k} = a_q \mathbf{W}_{k}$ with the binary property of $a_q, \forall q$. If $a_q=1$, constraint $\mbox{C11a}$ always holds because $\mathrm{Tr}\left( \mathbf{W}_{k} \right) \leq (Q-1)P_{\mathrm{max}}$. $\mbox{C11b}$ becomes $\overline{\mathbf{W}}_{q,k} \succeq \mathbf{W}_k$. Together with $\mbox{C11c}$, we have $\overline{\mathbf{W}}_{q,k} = \mathbf{W}_{k}$. If $a_q = 0$, constraint $\mbox{C11b}$ always holds. $\mbox{C11a}$ becomes $\overline{\mathbf{W}}_{q,k} \preceq \mathbf{0}$. Together with $\mbox{C11d}$, we have $\overline{\mathbf{W}}_{q,k} = \mathbf{0}$. Similarly, we introduce the constraints $\mbox{C12a-C12d}$ to decouple $a_q$ and $\mathbf{R}$ by replacing $\overline{\mathbf{W}}_{q,k}$ with $\overline{\mathbf{R}}_{q}$, and $\mathbf{W}_{k}$ with $\mathbf{R}$ in $\mbox{C11a-C11d}$. To decouple $a_q$ and $\mathbf{J}$, we introduce the following constraints 
\begin{eqnarray}
&&\mbox{C13a:}\hspace*{1mm}\overline{\mathbf{J}}_{q} \preceq a_{q} J_{\mathrm{max}} \mathbf{I},\hspace*{1mm}\forall q,\\
&&\mbox{C13b:}\hspace*{1mm}\overline{\mathbf{J}}_{q} \succeq \mathbf{J}-(1-a_{q}) J_{\mathrm{max}} \mathbf{I},\hspace*{1mm}\forall q,\\
&&\mbox{C13c:}\hspace*{1mm}\overline{\mathbf{J}}_{q} \preceq \mathbf{J},\hspace*{1mm}\forall q,\hspace*{2mm}\mbox{C13d:}\hspace*{1mm}\overline{\mathbf{J}}_{q} \succeq \mathbf{0},\hspace*{1mm}\forall q, 
\end{eqnarray}
where $J_{\mathrm{max}}$ is a pre-determined value to make sure that $\mathrm{Tr} \left( \mathbf{J} \right) \leq J_{\mathrm{max}}$. To handle the coupling between $a_q$ and $\epsilon_{ij}^{(q)}$, we introduce new optimization variables $\overline{\epsilon}_{ij}^{(q)} \geq 0, \forall i$, $j$, to replace $(1-a_q) \epsilon_{ij}^{(q)}$ in constraint $\myoverline{\mbox{C10}}$. With the obtained $\overline{\epsilon}_{ij}^{(q)}$ after solving the problem, we can recover $\epsilon_{ij}^{(q)}$ as follows. If $a_q = 1$, then $\epsilon_{ij}^{(q)} = 0$, and if $a_q = 0$, $\epsilon_{ij}^{(q)} = \overline{\epsilon}_{ij}^{(q)}$.
Then, the optimization problem is equivalently transformed to
\begin{eqnarray}
\label{big_M}
&&\hspace{-9mm}\underset{\substack{ \mathbf{W}_k, \mathbf{R}, \mathbf{J}, a_q,\\ \overline{\mathbf{W}}_{q,k}, \overline{\mathbf{R}}_q, \overline{\mathbf{J}}_q, \overline{\epsilon}_{ij}^{(q)} }}{\mino} \hspace{2mm} \mathrm{Tr}\left( \mathbf{J}^{-1} \right) \notag\\
&&\hspace*{-7mm}\subto\hspace*{4mm} \overline{\mbox{C1}}, \overline{\mbox{C2}}, \mbox{C3}, \mbox{C4}, \mbox{C8}, \mbox{C9}, \myoverline{\mbox{C10}}, \notag \\
&&\hspace*{14mm} \mbox{C11a-C11d}, \mbox{C12a-C12d}, \mbox{C13a-C13d}.
\end{eqnarray}
\begin{remark}
Note that we choose big-M instead of alternating optimization (AO) to deal with the coupling in $a_q \mathbf{W}_k$ because the solution of $a_q$ tends to be trapped into the first round of AO due to the binary property of $a_q$. As a result, the solution of the other optimization variables is trapped accordingly with the given $a_q$. This makes the algorithm more susceptible to being trapped in an unsatisfactory stationary point. Different from AO, the big-M method keeps the structure of the coupling relations by introducing the auxiliary variable $\overline{\mathbf{W}}_{q,k}$ and the constraints C11a-C11d.
\end{remark}
\par
To handle the binary constraint C4 on the AP selection variables $a_q, q \in \mathcal{Q}$, we rewrite constraint C4 equivalently as
\begin{eqnarray}
\mbox{C4a:} \hspace*{1mm} \underset{q\in\mathcal{Q}}{\sum}(a_{q}-a_{q}^2)\leq 0, \hspace*{4mm}
\mbox{C4b:} \hspace*{1mm} 0\leq a_{q} \leq 1,\hspace*{1mm}\forall q,
\end{eqnarray}
where constraint C4a is still non-convex. To overcome this issue, we exploit the penalty method \cite{nocedal1999numerical} and introduce a penalty term in the objective function of problem \eqref{big_M} as follows
\begin{eqnarray}
\label{penalty}
&&\hspace{-9mm}\underset{\substack{ \mathbf{W}_k, \mathbf{R}, \mathbf{J}, a_q,\\ \overline{\mathbf{W}}_{q,k}, \overline{\mathbf{R}}_q, \overline{\mathbf{J}}_q, \overline{\epsilon}_{ij}^{(q)} }}{\mino} \hspace{2mm} f_1 \overset{\triangle}{=} \mathrm{Tr}\left( \mathbf{J}^{-1} \right) + \mu \underset{q\in\mathcal{Q}}{\sum}\left(-a_{q}^2 + a_{q} \right) \notag\\
&&\hspace*{-7mm}\subto\hspace*{4mm} \overline{\mbox{C1}}, \overline{\mbox{C2}}, \mbox{C3}, \mbox{C4b}, \mbox{C8}, \mbox{C9}, \myoverline{\mbox{C10}}, \notag \\
&&\hspace*{14mm} \mbox{C11a-C11d}, \mbox{C12a-C12d}, \mbox{C13a-C13d}.
\end{eqnarray}
where $\mu > 0$ is the penalty factor to force the solution to satisfy constraint C4a. The equivalence between problem \eqref{big_M} and problem \eqref{penalty} is revealed in the following theorem.
\par
\textbf{Theorem 1}:\hspace*{1mm} Denote $a_q^{(r)}$ as the optimal solution of problem \eqref{penalty} for penalty factor $\mu = \mu_r$. When $\mu_r$ is sufficiently large, i.e., $\mu_r \rightarrow \infty$, every limit point $\bar{a}_q$ of the sequence $\left\{ a_q^{(r)} \right\}$ is an optimal solution of problem \eqref{big_M}.
\par
\textit{Proof:}\hspace*{1mm}\textit{Theorem 1} can be proved by similar steps as in \cite[Appendix B]{9183907}. The detailed
proof is omitted due to space limitation. \QED

Then, we employ SCA to tackle the non-convexity of the penalty term. Specifically, we construct a convex surrogate function for the objective function $f_1$ by applying the first-order Taylor expansion on the term $-a_q^2$. Given the obtained solution in the $(t-1)$-th iteration, the optimization problem to be solved in the $t$-th iteration is written as
\begin{eqnarray}
\label{SCA}
&&\hspace{-9mm}\underset{\substack{ \mathbf{W}_k, \mathbf{R}, \mathbf{J}, a_q,\\ \overline{\mathbf{W}}_{q,k}, \overline{\mathbf{R}}_q, \overline{\mathbf{J}}_q, \overline{\epsilon}_{ij}^{(q)} }}{\mino} \hspace{2mm} \overline{f}_1^{(t)} \notag\\
&&\hspace*{-7mm}\subto\hspace*{4mm} \overline{\mbox{C1}}, \overline{\mbox{C2}}, \mbox{C3}, \mbox{C4b}, \mbox{C8}, \mbox{C9}, \myoverline{\mbox{C10}}, \notag \\
&&\hspace*{14mm} \mbox{C11a-C11d}, \mbox{C12a-C12d}, \mbox{C13a-C13d},
\end{eqnarray}
where the objective function at the $t$-th iteration is given by
\begin{eqnarray}
\overline{f}_1^{(t)} \overset{\triangle}{=} \mathrm{Tr}\left( \mathbf{J}^{-1} \right)\hspace*{-0.5mm}+ \hspace*{-0.5mm}\mu\underset{q\in\mathcal{Q}}{\sum}\hspace*{-0.5mm}\left( \big(1-2a^{(t-1)}_{q} \big)a_{q} \hspace*{-0.5mm}+\hspace*{-0.5mm} \big(a^{(t-1)}_{q} \big)^2 \right),
\end{eqnarray}
which is a convex function with respect to $\mathbf{J}$ and $a_q, \forall q$.
\par
Up to now, the only non-convexity of \eqref{SCA} stems from the rank-one constraint C8: $\mathrm{Rank}\left( \mathbf{W}_k \right) = 1$. To tackle this problem, we relax the rank-one constraint by applying the SDR technique. The relaxed problem, denoted as (SDR1), is a convex problem and can be solved efficiently by standard convex optimization
solvers such as CVX \cite{grant2014cvx}. The tightness of the rank relaxation is presented in the following proposition.
\par
\begin{prop}
Based on the obtained solution $\left \{  \mathbf{W}_k^* \right \} $, $\mathbf{R}^*$, $\mathbf{J}^*$, $\overline{\mathbf{W}}_{q,k}^*$, $\overline{\mathbf{R}}_q^*$, $\overline{\mathbf{J}}_q^*$, $a_q^*$ and $({\overline{\epsilon}_{ij}^{(q)}})^*$ to (SDR1), there exits the equivalent solution $\left \{  \widetilde{\mathbf{W}}_k^* \right \} $, $\widetilde{\mathbf{R}}^*$, $\widetilde{\mathbf{J}}^*$, $\widetilde{\overline{\mathbf{W}}}_{q,k}^*$, $\widetilde{\overline{\mathbf{R}}}_q^*$, $\widetilde{\overline{\mathbf{J}}}_q^*$, $\widetilde{a}_q^*$ and $(\widetilde{\overline{\epsilon}}_i^{(q)})^*$ that achieves the same objective value and satisfies $\mathrm{Rank} \left \{ \widetilde{\mathbf{W}}_k^* \right \} = 1, \forall k.$
\end{prop}
\par
\textit{Proof:} Please refer to the Appendix B. \QED
\par
The penalty-based algorithm for solving \eqref{penalty} is summarized in $\textbf{Algorithm 1}$. Note that the objective value of \eqref{penalty} is upper bounded by the minimum of (SDR1). By iteratively solving problem (SDR1) in $\textbf{Algorithm 1}$, we can monotonically tighten the upper bound and obtain a sequence of solutions. The objective values of (SDR1) achieved by the sequence of the solutions are monotonically non-increasing and the proposed algorithm is guaranteed to converge to a stationary point of \eqref{penalty} in polynomial time \cite{dinh2010local}. According to \cite[Theorem 3.12]{bomze2010interior}, the computational complexity of $\textbf{Algorithm 1}$ is given by $\mathcal{O}\left( \log \left( \frac{1}{\varepsilon_1} \right) Q^4(2N+1)^3 + Q^4KN^3 + Q^4K^2N^2 +  Q^3K^3 \right)$.



\begin{algorithm}[t]
\caption{Penalty-based algorithm for sensing phase}
\begin{algorithmic}[1]
\small
\STATE Set iteration index $t=1$, error tolerance factor $0<\varepsilon_1\ll1$, and penalty factor $\mu$. Initialize the AP selection variable $a_q = 0, \forall q$
\REPEAT
\STATE Solve problem (SDR1) for fixed $a_q^{(t-1)}, \forall q$
\STATE Update $a_q^{(t)}$
\STATE Set $t=t+1$
\UNTIL $\frac{\left|\overline{f}^{(t)}_1-\overline{f}_1^{(t-1)}\right|}{\overline{f}_1^{(t)}}\leq \varepsilon_1$
\end{algorithmic}
\end{algorithm}

\subsection{Solution of Problem \eqref{step2}}
After solving problem \eqref{step1}, we obtain $\Omega^+$, $a_q$, and $P_{\mathrm{EH}}^{\mathrm{I}}$ which are used for solving problem \eqref{step2}. The non-linear energy harvesting model is first transformed into several constraints. Then, we use S-procedure to handle the robust WPT constraint. Fractional programming is employed to deal with the fractional form of the constraint that models the relation between the input power and the harvested power at the EHR.
\par
Similar to the steps of solving problem \eqref{step1}, we introduce the beamforming matrices $\mathbf{W}_k,\hspace{1mm} \forall k$, resulting in the rank-one constraint C14: $\mathrm{Rank}\left( \mathbf{W}_k \right) = 1$.
We use the same transformation as $\overline{\mbox{C1}}$ and $\overline{\mbox{C2}}$ in problem \eqref{step1} to equivalently reformulate C5 and C6
into $\overline{\mbox{C5}}$ and $\overline{\mbox{C6}}$, respectively. The difficulty of solving problem \eqref{step2} lies in the robust design of the nonlinear energy harvesting model in constraint $\mbox{C7}$. To handle this, we first introduce two auxiliary optimization variables $P_{\mathrm{IN}}$ and $P_{\mathrm{EH}}$, and equivalently rewrite constraint C7 into the following constraints
\begin{eqnarray}
&&\hspace{-4mm}\mbox{C7a:} \hspace{1mm} \underset{\Delta \mathbf{p}^+ \in \Omega^+ }{\min} \underset{k\in\mathcal{K}}{\sum} \underset{q \in \mathcal{Q}}{\sum} (1-a_q) \mathrm{Tr} \left (  \mathbf{A}_q \mathbf{H}_{\text{EH}} \mathbf{A}_q \mathbf{W}_k \right ) \notag \\
&&\hspace{10mm}+  \underset{q \in \mathcal{Q}}{\sum} (1-a_q) \mathrm{Tr} \left (  \mathbf{A}_q \mathbf{H}_{\text{EH}} \mathbf{A}_q \mathbf{R} \right )
\geq P_{\mathrm{IN}}, \\
&&\hspace{-4mm}\mbox{C7b:} \hspace{1mm} \frac{1}{1+\exp{ \left [ -a \left ( P_{\mathrm{IN}}-b \right ) \right ] }} \geq (1-\Xi) \frac{P_{\mathrm{EH}}}{P_{\mathrm{sat}}} + \Xi, \\
&&\hspace{-4mm}\mbox{C7c:} \hspace{1mm} \eta P_{\mathrm{EH}}^{\mathrm{I}} + (1-\eta) P_{\mathrm{EH}} \geq P_{\mathrm{req}},
\end{eqnarray}
where $\mathbf{H}_{\text{EH}} \overset{\triangle}{=} \mathbf{h}_{\text{EH}} \mathbf{h}_{\text{EH}}^H $.
The uncertainty in constraint $\mbox{C7a}$ comes from the geometry information of APs and EHR in $\mathbf{h}_{\mathrm{EH}}^H$. To facilitate the robust design, we use the first-order Taylor expansion to
approximate $\mathbf{h}_{\mathrm{EH}}^H$ as follows
\begin{eqnarray}
\mathbf{h}_{\mathrm{EH}} = \overline{\mathbf{h}}_{\mathrm{EH}} + \dot{\mathbf{h}}_{\mathrm{EH},1} \Delta p_1^+ + \dot{\mathbf{h}}_{\mathrm{EH},2} \Delta p_2^+,
\end{eqnarray}
where $\overline{\mathbf{h}}_{\mathrm{EH}}$ is the estimated value of $\mathbf{h}_{\mathrm{EH}}$ at $\overline{\mathbf{p}}$. 
$\dot{\mathbf{h}}_{\mathrm{EH},1}$ and $\dot{\mathbf{h}}_{\mathrm{EH},2}$ denote the derivatives of $\mathbf{h}_{\mathrm{EH}}$ with respect to $p_1$ and $p_2$, respectively. For notational convenience, 
we define $\mathbf{E} = \left[ \dot{\mathbf{h}}_{\mathrm{EH},1}, \hspace*{1mm}\dot{\mathbf{h}}_{\mathrm{EH},2} \right] \in \mathbb{C}^{NQ \times 2}$. Then, $\mathbf{h}_{\mathrm{EH}}$ is rewritten as 
\begin{eqnarray} \label{taylor_2}
\mathbf{h}_{\mathrm{EH}} = \overline{\mathbf{h}}_{\mathrm{EH}} + \mathbf{E} \Delta \mathbf{p}^+.
\end{eqnarray}
Next, we tackle the robust design in constraint C7a by employing the S-procedure lemma, which is described in the following.
\begin{lemma}
\textit{(S-Procedure \cite{boyd2004convex})} Two functions $f_i(\mathbf{t}): \mathbb{C}^{N\times 1}\to \mathbb{R}$, $i\in \left \{ 1,2 \right \}$ are defined as
\begin{equation}
f_i(\mathbf{t})= \mathbf{t}^H\mathbf{A}_i\mathbf{t}+2\Re\left \{\mathbf{b}^H_i\mathbf{t}  \right \} + c_i,
\end{equation}
where $\mathbf{A}_i\in \mathbb{H}^N$, $\mathbf{b}_i\in \mathbb{C}^{N\times 1}$, and $\mathrm{c}_i\in \mathbb{R}$. Then, the implication $f_1(\mathbf{t})\leq0 \Rightarrow f_2(\mathbf{t})\leq0$ holds if and only if there exists a variable $\kappa \geq 0$ such that
\begin{equation}
\kappa 
\begin{bmatrix}
\mathbf{A}_1 &  \mathbf{b}_1\\
\mathbf{b}_1^H &  \mathit{c}_1
\end{bmatrix}-\begin{bmatrix}
\mathbf{A}_2 &  \mathbf{b}_2\\
\mathbf{b}_2^H &  \mathit{c}_2
\end{bmatrix}\succeq \mathbf{0}.
\end{equation}
\end{lemma}
\par
To facilitate the application of S-procedure, we substitute \eqref{taylor_2} into constraint C7a, and rewrite constraint C7a as
\begin{eqnarray}
&& \hspace{-8mm} 
\mbox{C7a}\mbox{:}
-(\Delta \mathbf{p}^+)^H \mathbf{E}^H \overline{\mathbf{S}}  \mathbf{E} \Delta \mathbf{p}^+ - 2\Re \left \{ \Bar{\mathbf{h}}^H   \overline{\mathbf{S}} \mathbf{E} \Delta \mathbf{p}^+ \right \} -  \Bar{\mathbf{h}}^H \overline{\mathbf{S}} \Bar{\mathbf{h}} \notag \\
&& + P_{\text{IN}} \leq 0, \hspace{1mm} \Delta \mathbf{p}^+ \in \Omega^+ ,
\end{eqnarray}
where $\overline{\mathbf{S}}$ is defined as
\begin{eqnarray}
\overline{\mathbf{S}} \overset{\triangle}{=}  \underset{q \in \mathcal{Q}}{\sum} (1-a_q) \mathbf{A}_q \left( \underset{k \in \mathcal{K}}{\sum} \mathbf{W}_k + \mathbf{R} \right) \mathbf{A}_q.
\end{eqnarray}
\par
We introduce the new optimization variable $\kappa \geq 0$ and apply \textbf{Lemma 2}. Then, the following implication 
\begin{eqnarray}
(\Delta \mathbf{p}^+)^H \Delta \mathbf{p}^+ - \left( (\psi_1^{\mathrm{+}})^2 + (\psi_2^{\mathrm{+}})^2 \right) \leq 0 \Rightarrow \mbox{C7a}
\end{eqnarray}
holds if and only if there exists $\kappa \geq 0$ satisfying the constraint $\overline{\mbox{C7a}}$ given by
\begin{eqnarray}
\overline{\mbox{C7a}}\mbox{:}
\begin{bmatrix}
\kappa \mathbf{I}_{3} & \hspace*{-0.5mm}\mathbf{0} \\
\mathbf{0} & \hspace*{-0.5mm}-\kappa \left( (\psi_1^{\mathrm{+}})^2 \hspace*{-0.5mm}+\hspace*{-0.5mm} (\psi_2^{\mathrm{+}})^2 \right) - P_{\text{IN}}
\end{bmatrix} + \mathbf{U}^H \overline{\mathbf{S}} \mathbf{U} \succeq \mathbf{0},
\end{eqnarray}
where $\mathbf{U} = \left[ \mathbf{E},\hspace*{1mm} \overline{\mathbf{h}}_{\text{EH}} \right]$.
Then, the optimization problem \eqref{step2} is transformed into
\begin{eqnarray}
\label{before_fp}
&&\hspace{-10mm}\underset{\substack{ \mathbf{W}_k, \mathbf{R}, \\ P_{\mathrm{IN}}, P_{\mathrm{EH}}, \kappa \geq 0}}{\mino} \hspace{4mm} f_2 \overset{\triangle}{=} P_{\mathrm{avg}} \notag\\
&&\hspace*{-9mm}\subto\hspace*{4mm} \overline{\mbox{C5}}, \overline{\mbox{C6}}, \overline{\mbox{C7a}}, \mbox{C7b}, \mbox{C7c}, \mbox{C14}.
\end{eqnarray}
\par
To tackle the non-convex constraint $\mbox{C7b}$, we employ the fractional programming method \cite{8314727}, which uses the quadratic transformation to recast the non-convex fractional form problem as a sequence of convex problems. In particular, we introduce an auxiliary variable $\varpi$ and transform constraint C7b as follows
\begin{eqnarray}
\overline{\mbox{C7b}}\mbox{:}\hspace*{1mm}2\varpi\hspace*{-0.5mm}-\hspace*{-0.5mm}\varpi^2\hspace*{-0.5mm}\Big( 1\hspace*{-1mm}+\hspace*{-0.5mm}\exp \big[a( b\hspace*{-0.5mm}-\hspace*{-0.5mm}P_{\mathrm{IN}})\big ] \Big)\hspace*{-0.5mm}\geq\hspace*{-0.5mm}\frac{(1\hspace*{-0.5mm}-\hspace*{-0.5mm}\Xi) P_{\mathrm{EH}}}{P_{\mathrm{sat}}}\hspace*{-0.5mm}+\hspace*{-0.5mm}\Xi.
\end{eqnarray}
Variable $\varpi$ and the other variables are optimized alternatingly. By fixing the other optimization variables, $\varpi$ is updated by
\begin{eqnarray} \label{varpi_update}
\varpi = \frac{1}{1+\exp{ \left [ a \left ( b -P_{\mathrm{IN}}\right ) \right ] }}.
\end{eqnarray}
On the other hand, for fixed $\varpi$, we solve the following problem
\begin{eqnarray}
\label{after_fp}
&&\hspace{-10mm}\underset{\substack{ \mathbf{W}_k, \mathbf{R}, \\ P_{\mathrm{IN}}, P_{\mathrm{EH}}, \kappa \geq 0}}{\mino} \hspace{3mm} f_2 = P_{\mathrm{avg}} \notag\\
&&\hspace*{-9mm}\subto\hspace*{4mm} \overline{\mbox{C5}}, \overline{\mbox{C6}}, \overline{\mbox{C7a}}, \overline{\mbox{C7b}}, \mbox{C7c}, \mbox{C14}.
\end{eqnarray}
Up to now, the only non-convexity of \eqref{after_fp} comes from the constraint C14: $\mathrm{Rank}\left( \mathbf{W}_k \right) = 1$. Again, we use the SDR relaxation by removing the rank-one constraint C14 and get the relaxed problem (SDR2), which is a convex problem and can be solved by standard convex
optimization solvers. The tightness of SDR relaxation is guaranteed via the following proposition.
\par
\begin{prop}
Based on the solution $\left \{ \mathbf{W}_k^* \right \} $, $\mathbf{R}^*$, $P_{\mathrm{IN}}^*$, $P_{\mathrm{EH}}^*$, and $\kappa^*$ of (SDR2), we can always construct the equivalent solution $\left \{  \widetilde{\mathbf{W}}_k^* \right \} $, $\widetilde{\mathbf{R}}^*$, $\widetilde{P}_{\mathrm{IN}}^*$, $\widetilde{P}_{\mathrm{EH}}^*$, and $\widetilde{\kappa}^*$ expressed as follows, which achieves the same objective value and satisfies $\mathrm{Rank} \left \{ \widetilde{\mathbf{W}}_k^* \right \} = 1, \forall k.$
\begin{eqnarray} \label{SDR_transformation2}
&&\hspace{-4mm}\widetilde{\mathbf{W}}_k^* = \frac{\mathbf{W}_k^* \mathbf{A}_{q_0} \mathbf{h}_k \mathbf{h}_k^H \mathbf{A}_{q_0} \mathbf{W}_k^*}{\mathbf{h}_k^H \mathbf{A}_{q_0} \mathbf{W}_k^* \mathbf{A}_{q_0} \mathbf{h}_k}, \label{W2} \\ 
&&\hspace{-4mm}\widetilde{\mathbf{R}}^* = \sum_{k=1}^K \mathbf{W}_k^* + \mathbf{R}^* - \sum_{k=1}^K \widetilde{\mathbf{W}}_k^*, \label{R2} \\
&&\hspace{-4mm}\widetilde{P}_{\mathrm{IN}}^* = P_{\mathrm{IN}}^*, \hspace{2mm} \widetilde{P}_{\mathrm{EH}}^* = P_{\mathrm{EH}}^*, \hspace{2mm} \widetilde{\kappa}^* = \kappa^*, \label{sdr_ja2}
\end{eqnarray}
where $q_0$ is defined in Appendix B.
\end{prop}
\par
\textit{Proof:} The proof is similar to that for $\textbf{proposition 2}$ and is omitted for simplicity. \QED
\par
The iterative fractional programming-based algorithm for solving problem \eqref{step2} is summarized in $\textbf{Algorithm 2}$. Some additional remarks on the algorithm are as follows:
\par
\textit{i) Initialization:} A set of feasible solutions is required to
initialize the problem (SDR2). However, some prior information regarding the problem (SDR2) such as the received power at the EHR is missing. In fact, a small value of $P_{\text{IN}}$ may cause infeasibility in solving the problem due to failure to meet the EHR's energy harvesting requirement. Furthermore, the maximum energy the system can allocate to the EHR is also unknown. To this end, we propose to initialize $\textbf{Algorithm 2}$ by maximizing the received power at the EHR, which is formulated as
\begin{align}
\label{intialize_fp}
&\underset{\substack{\mathbf{W}_k, \mathbf{R}, \\ P_{\mathrm{IN}} \geq 0}} {\maxo} \hspace{4mm} P_{\mathrm{IN}} \notag\\
&\hspace*{-3mm}\subto\hspace*{2mm} \overline{\mbox{C5}}, \overline{\mbox{C6}}, \overline{\mbox{C7a}}.
\end{align}
In particular, by comparing the $P_{\text{IN}}$ obtained from \eqref{intialize_fp} with the minimum required received power at the EHR, which can be calculated based on $P_{\text{req}}$ and the energy harvesting model, we can check the feasibility. Specifically, if the obtained $P_{\text{IN}}$ is smaller than the minimum required received power at the EHR, the problem (SDR2) is verified to be infeasible.
\par
\textit{ii) Convergence and Complexity:} According to \cite{8314727}, the algorithm is guaranteed to converge to a stationary point of problem \eqref{before_fp} in polynomial time.
The computational complexity of the $\textbf{Algorithm 2}$ is dominated by step 4 \cite{arora2009computational}. The computational complexity is given by $\mathcal{O}\left( \log (\frac{1}{\varepsilon_2}) (Q+K)N^3(Q-1)^3 + (Q+K)^2N^2(Q-1)^2 + ( \right.$ $\left. Q+K)^3 \right)$ \cite[Theorem 3.12]{bomze2010interior}.
{\color{black}\begin{algorithm}[t]
\caption{Fractional programming-based algorithm for WPT phase}
\begin{algorithmic}[1]
\small
\STATE Set iteration index $t=1$, error tolerance factor $0<\varepsilon_2 \ll1$. Initialize the optimization variables $\mathbf{W}_k$, $\mathbf{R}$, $P_{\mathrm{IN}}$ by solving the problem \eqref{intialize_fp}
\REPEAT
\STATE Update $\varpi$ by \eqref{varpi_update}
\STATE Update $\mathbf{W}_k$ and $\mathbf{R}$ by solving problem (SDR2) with obtained $\varpi$
\STATE Set $t=t+1$
\UNTIL $\frac{\left|f_2^{(t)}-f_2^{(t-1)}\right|}{f_2^{(t)}}\leq \varepsilon_2$
\end{algorithmic}
\end{algorithm}}
\subsection{Solution of the Time Allocation Strategy}
We use line search to traverse the time-splitting ratio $\eta$ for two phases. Specifically, we denote the initial value and step increment of $\eta$ by $\eta_0$ and $\Delta \eta$, respectively. The time-splitting ratio at the $d$-th iteration of the outer layer is $\eta = \eta_0 + (d-1) \Delta \eta$, where $d \in \mathbb{N}^+$. The search area is $\Upsilon = \left\{ \eta_0 + d \Delta \eta \mid \eta_0 \leq \eta_0 + d \Delta \eta \leq 1, d \in \mathbb{N}^+ \right\}$. In each iteration of the outer layer, we solve the problems (SDR1) and (SDR2) alternatingly and check the objective value of (SDR2) for each given time-splitting policy. The optimal $\eta^*$ is selected as the one associated with the maximum objective function value of (SDR2), i.e., $\eta^* = \underset{\eta \in \Upsilon}{\maxo} \hspace{1mm} f_2.$
The overall two-layer algorithm is summarized in $\textbf{Algorithm 3}$.
\begin{algorithm}[t]
\caption{{\color{black}Two-layer algorithm}}
\begin{algorithmic}[1]
\small
\STATE Set $\eta_0$, $\Delta \eta$, and iteration index $d = 1$
\REPEAT 
\STATE Generate the time-splitting policy $\eta = \eta_0 + (d-1) \Delta \eta$
\STATE Solve problem (SDR1) for given $\eta$ by $\textbf{Algorithm 1}$ and obtain the updated location uncertainty $\psi^{+}_i, i \in \{ 1,2 \}$ and the AP selection variables $a_q, \forall q$
\STATE Based on the obtained $\psi^{+}_i$ and $a_q$, solve problem (SDR2) by $\textbf{Algorithm 2}$
\STATE Store the objective value $f_2$ and the corresponding solutions
\STATE Set $d=d+1$
\UNTIL $\eta_0 + d \Delta \eta \geq 1$
\STATE Select the solution corresponds to the maximum value of $f_2$
\end{algorithmic}
\end{algorithm}

\section{Numerical Results}
This section utilizes numerical results to validate the effectiveness of the proposed algorithm. In the simulation, we consider a SWIPT system consisting of $Q=3$ APs, one EHR, and $K=5$ CUs. Each AP is equipped with a ULA with $N=6$ antennas. The Cartesian coordinates of three APs are $[-15,15]$, $[15,15]$ and $[0,-15]$, where the unit is meter. The EHR and CUs are randomly and uniformly distributed in the area $\left\{ (x,y) \mid -20 \leq x \leq 20, -15 \leq y \leq 15 \right\}$.
The parameters of the energy harvesting model are set as $a = 7400$, $b = 0.0001$, and $P_{\text{sat}}=0.0003$ \cite{7264986}.
The path loss at the reference distance of 1 m is set to 40 dB and the distance-dependent path loss exponents of the AP-CU and AP-EHR links are set to 3 and 2, respectively \cite{ref1}. For the small-scale fading, we consider Rayleigh channel models. Without loss of generality, we assume that the SINR requirements of CUs are identical, i.e., $\Gamma_k=\Gamma, \forall k \in \mathcal{K}$ and the location uncertainties are also identical for two coordinates, i.e., $\psi_i=\psi, i=1,2$. Unless otherwise specified, we adopt the parameters in Table I.
\begin{table}[t]\vspace*{0mm}\caption{System Simulation Parameters.\vspace*{0mm}}\label{parameter}\footnotesize
\newcommand{\tabincell}[2]{\begin{tabular}{@{}#1@{}}#2\end{tabular}}
\centering
\renewcommand{\arraystretch}{1.2}
\begin{tabular}{|l|l|l|}
\hline
    \hspace*{-1mm}$\eta_0$ & Initial time-splitting ratio value & $0.1$ \\
\hline
    \hspace*{-1mm}$ \Delta \eta$ & Step increment of time-splitting ratio & $0.1$ \\
\hline
    \hspace*{-1mm}$ L$ & Number of time slots in each frame & $1024$ \cite{10364735} \\
\hline
    \hspace*{-1mm}$P_{\mathrm{max}}$ & Maximum transmit power at each AP & $3$ W \\
\hline
    \hspace*{-1mm}$\Gamma$ &  Minimum SINR requirement & $12$ dB \\
\hline
    \hspace*{-1mm}$P_{\mathrm{req}}^{\mathrm{EH}}$ & Minimum energy harvesting requirement & $0.08$ mW \cite{10382465}\\
\hline
    \hspace*{-1mm}$\psi$ & Initial location uncertainty& $1.2$ m \\
\hline
    \hspace*{-1mm}$\varepsilon_1, \varepsilon_2$ & Error tolerance factors & $10^{-3}$ \\
\hline
    \hspace*{-1mm}$\sigma_k^2$ & Noise power at the $k$-th CU & $-60$ dBm \\
\hline
    \hspace*{-1mm}$\sigma_r^2$ & Noise power at the sensing receiver & $-60$ dBm \\
\hline
\end{tabular}
\end{table}
\subsection{Baseline Schemes}
For comparison purposes, we consider the following four baseline schemes.
\par
\textbf{Baseline scheme 1:} Baseline scheme 1 adopts the traditional robust design without the sensing phase to handle the location uncertainty. It minimizes power consumption while 
satisfying the QoS of the CUs and the EHR with the initial location error by optimizing the beamforming policy.
\par
\textbf{Baseline scheme 2:} 
With the two-phase design, baseline scheme 2 employs the robust zero-forcing (ZF) beamforming design where the multiuser interference and the interference from radar signals to CUs are eliminated.
In particular, the beamforming policy $\mathbf{W}=\left[ \mathbf{w}_1, \cdots,  \mathbf{w}_K\right]$ and $\mathbf{R}$ satisfy 
\begin{eqnarray} \label{zf}
\mathbf{H}\mathbf{W} = \diag(\sqrt{\mathbf{p}}), \hspace{2mm}
\mathbf{H}\mathbf{R}=\mathbf{0},
\end{eqnarray}
where $\mathbf{H} \overset{\triangle}{=} \left[ \mathbf{h}_1,\cdots,\mathbf{h}_K \right]^H$ and $\mathbf{p}=\left[ 
p_1,\cdots,p_K \right]^T$ with $p_k$ denoting the signal power for the $k$-th CU. \eqref{zf} is equivalent to \cite{9124713}
\begin{eqnarray} \label{new_zf}
\mathbf{H} \left( \underset{k \in \mathcal{K}}{\sum}\mathbf{W}_k + \mathbf{R} \right) \mathbf{H}^H = \diag(\mathbf{p}).
\end{eqnarray}
The optimization problem for robust ZF design is formulated by introducing the new constraint \eqref{new_zf} and replacing the SINR constraints C2 and C6 in \eqref{step1} and \eqref{step2} with $p_k \geq \Gamma_k \sigma_k^2$.
\par
\textbf{Baseline scheme 3:} 
Baseline scheme 3 selects the sensing receiver randomly in the sensing phase, based on which robust WPT is then performed.
\par
\textbf{Baseline scheme 4:} Baseline scheme 4 employs the non-robust design algorithm which takes the estimated location as the true value. 
Without the robust design, Baseline scheme 4 consumes much lower power compared to the other baseline schemes. However, it may not be able to satisfy the requirements of EHR. To assess the practical performance of baseline scheme 4, we investigate whether its obtained solution satisfies the robust energy harvesting requirement of EHR by analyzing the infeasibility rate.
\begin{figure}[t]
\centering
\includegraphics[width=3.3in]{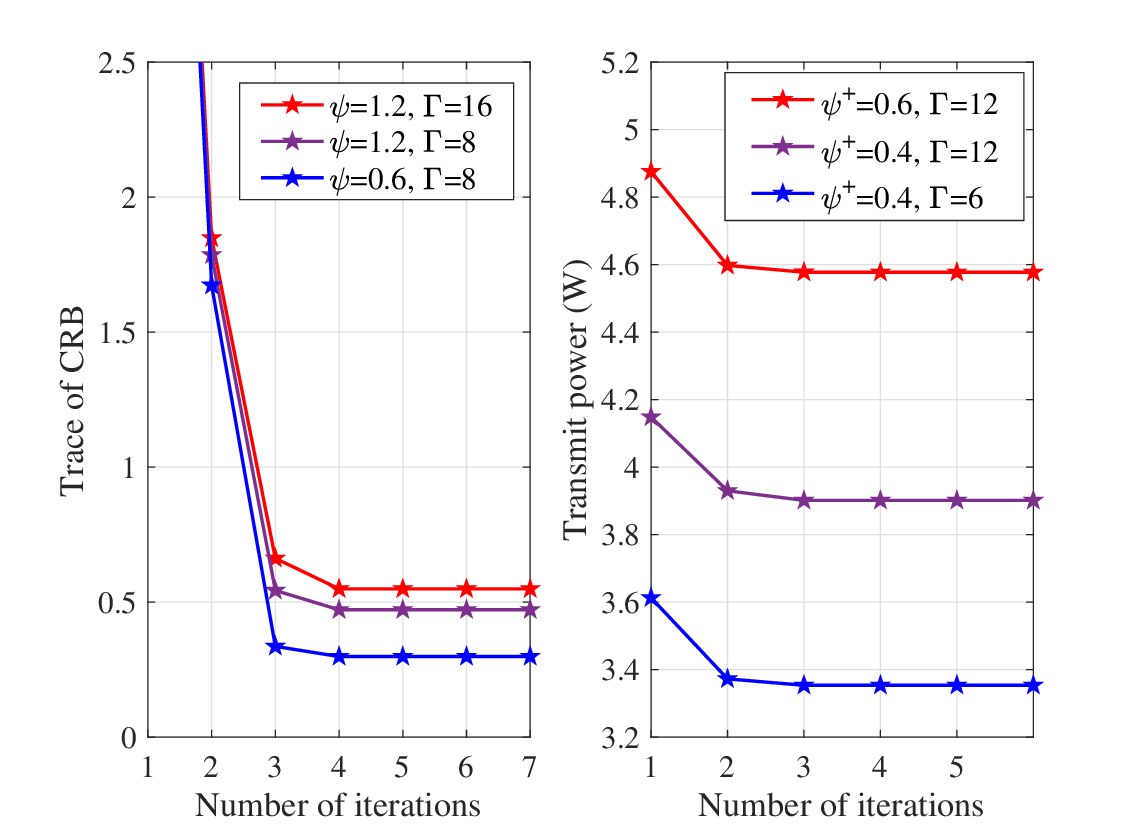}
\caption{Convergence of $\textbf{Algorithm 1}$ (Left) and $\textbf{Algorithm 2}$ (Right).}
\label{figure:convergence}
\end{figure}
\begin{figure}[t]
\centering
\includegraphics[width=3.3in]{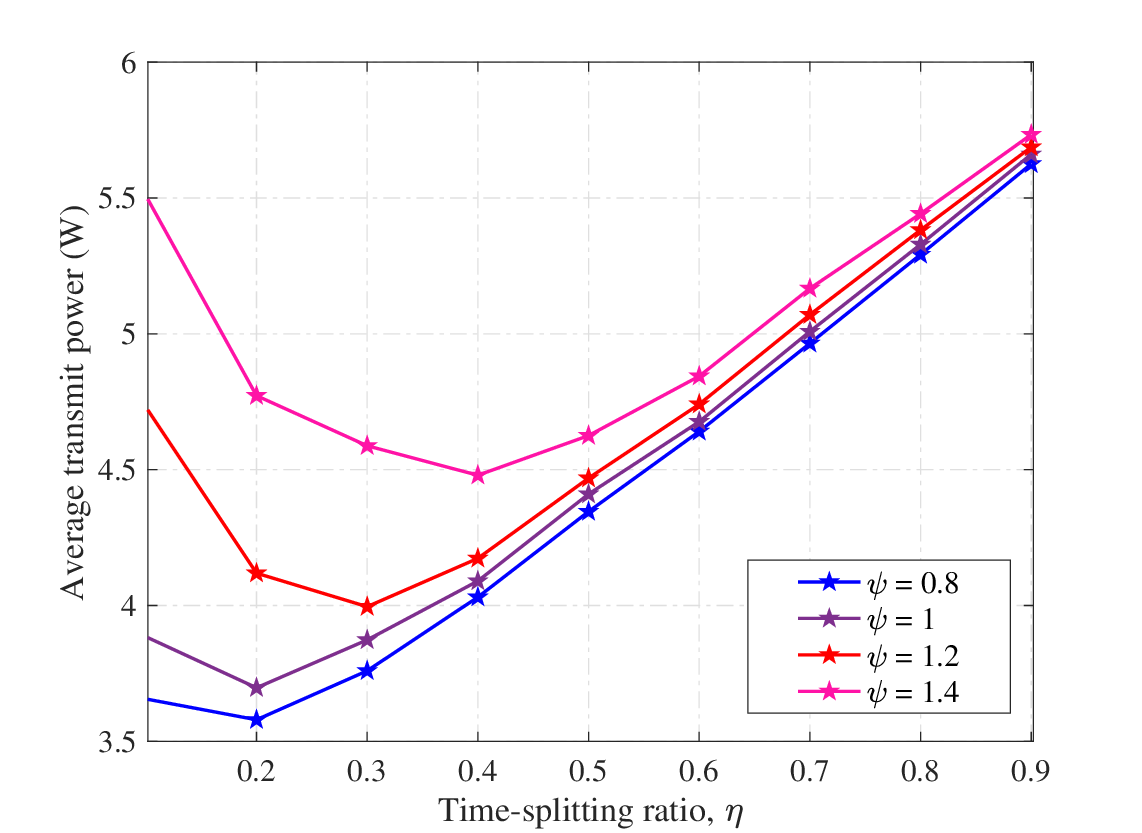}
\caption{Average transmit power versus time-splitting ratio.}
\label{figure:time_splitting}
\end{figure}
\begin{figure}[t]
\centering
\includegraphics[width=3.3in]{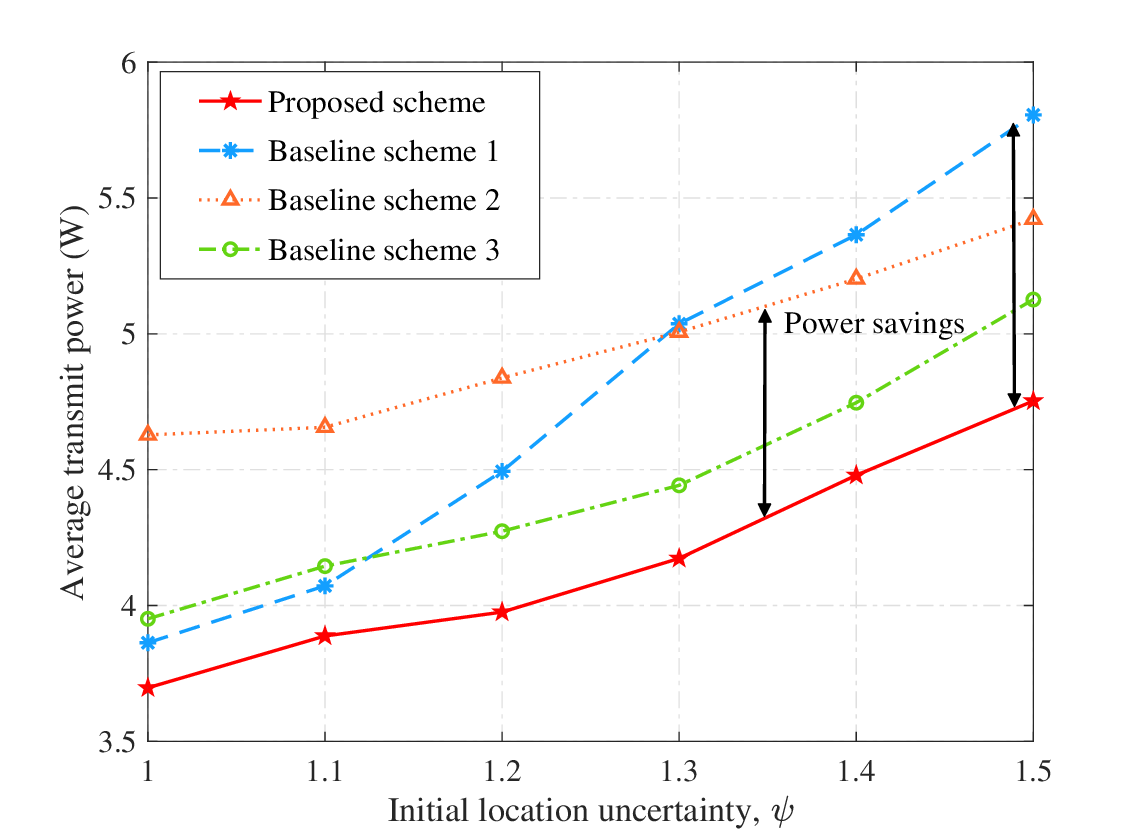}
\caption{Average transmit power versus initial location uncertainty.}
\label{figure:p_psi}
\end{figure}
\begin{figure}[t]
\centering
\includegraphics[width=3.3in]{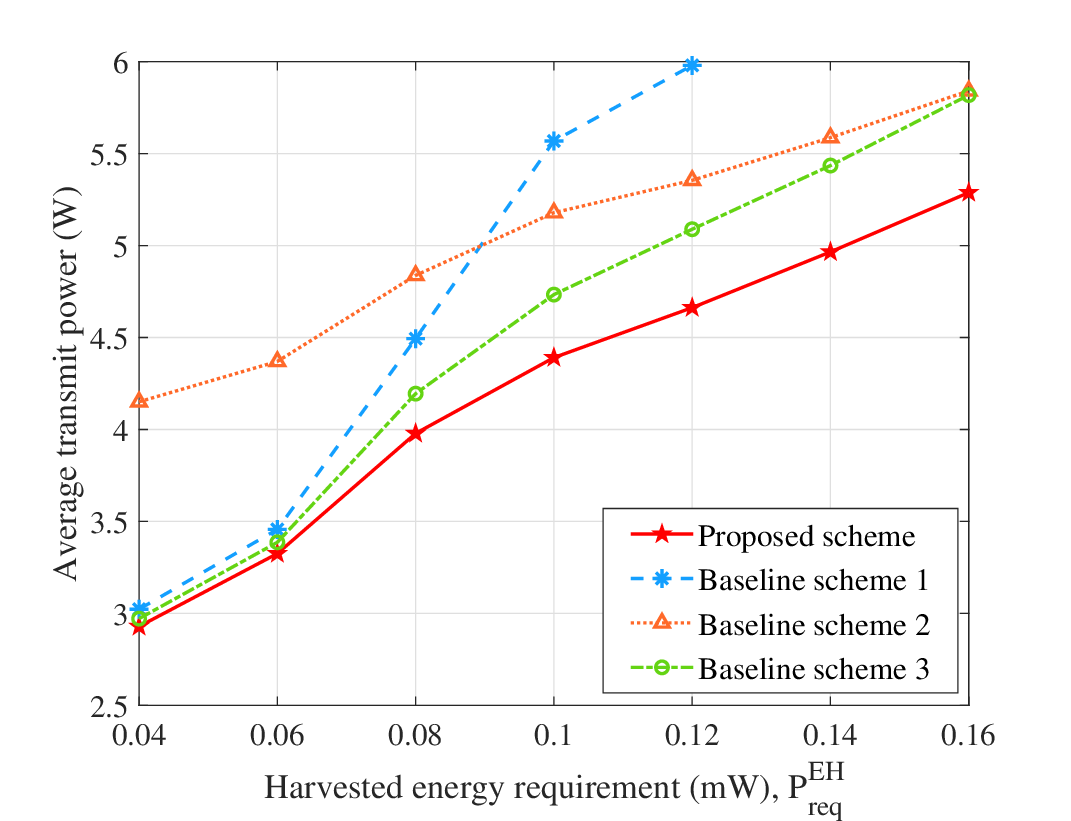}
\caption{Average transmit power versus harvested energy requirement.}
\label{figure:P_Peh}
\end{figure}
\subsection{Convergence of the Proposed Algorithm}
Fig. \ref{figure:convergence} illustrates the convergence of $\textbf{Algorithm 1}$ (left figure) and $\textbf{Algorithm 2}$ (right figure) with different location uncertainty and SINR requirements. It can be observed that 
both algorithms converge quickly. With increased location uncertainty and SINR requirements, the feasible sets of
both problems are reduced, causing performance degradation. In particular, for the sensing phase, the sensing performance decreases when the location uncertainty is large. The high SINR requirement of CUs takes more resources for communication and thus harms the sensing performance.
For the WPT phase, large location uncertainty and high SINR requirements cause more energy consumption.
\subsection{Time Allocation between Two Phases}
Fig. \ref{figure:time_splitting} shows the average power consumption of the system versus the time-splitting ratio for different initial location uncertainty $\psi$. It can be observed that, for each given $\psi$, there is an optimum splitting ratio. Take the case with $\psi = 1.2$ as an example. As $\eta$ increases from $0.1$ to $0.3$, more time is used to refine the localization accuracy.  This helps to achieve efficient WPT and precise beamforming thus saving the power. However, as $\eta$ further increases from $0.3$ to $0.9$, much more power is consumed in the sensing phase. As a result, the power saved in the WPT phase due to more accurate location information is  unable to compensate for the
power consumption in the sensing phase.
In addition, it is observed that a longer time should be allocated to the sensing phase given a larger initial location error.
\subsection{Effect of Initial Location Uncertainty}
Fig. \ref{figure:p_psi} presents the average power consumption versus the initial location uncertainty $\psi$. As expected, the power consumption increases monotonically with the initial location uncertainty and the proposed scheme achieves the minimum power consumption. 
The gap between the proposed scheme and baseline scheme 1 comes from the efficient WPT of the proposed scheme with refined location information.
In addition, the gap increases quickly with location uncertainty. This is because the proposed scheme benefits from the extra DoF brought by the time-splitting ratio and can adaptively determine the time allocation to the sensing phase for different initial location uncertainties.
Compared with baseline scheme 2, the proposed scheme jointly designs the beamforming vectors and the radar covariance matrix to fully exploit the DoF in beamforming design, which enlarges the feasible set of the optimization problem and hence leads to better system performance.
Different from baseline scheme 3, which
adopts the random AP selection, the proposed scheme benefits from the macro-diversity brought by the joint beamforming and AP selection design in the sensing phase, resulting in the sensing performance improvement. Hence, with smaller location uncertainty, the proposed scheme achieves more efficient WPT and saves power.
\par
Based on the above observation, we discuss the impact of the number of APs and CUs. More APs provide more diversity for sensing and powering the EHR, leading to improved sensing performance. Hence, a shorter sensing phase is able to achieve the optimal sensing performance. As a result, the optimal time-splitting ratio becomes smaller, i.e., less time is allocated to the sensing phase.
When the number of CUs increases, the requirements on the communication side increase. As a result, the sensing performance is worsened due to reduced resource allocated to sensing. Hence, it requires a longer time to achieve the same sensing performance and the optimal time-splitting ratio increases.
\subsection{Influence of Energy Harvesting Requirement}
Fig. \ref{figure:P_Peh} depicts the average power consumption with respect to the energy harvesting requirement $P_{\mathrm{req}}^{\mathrm{EH}}$. As shown in the figure, the power consumption of the system increases with $P_{\mathrm{req}}^{\mathrm{EH}}$ and the proposed scheme outperforms all baseline schemes. The gap between the proposed scheme and baseline schemes 1 and 3 is tiny when $P_{\mathrm{req}}^{\mathrm{EH}}$ is small. This is because the performance improvement of the proposed scheme comes from the robust WPT with better sensing results. However, for small $P_{\mathrm{req}}^{\mathrm{EH}}$, the robust WPT constraint is not that stringent, and thus the performance improvement is limited. In addition, it can be observed that the power consumption of baseline scheme 1 becomes very large when the energy harvesting requirement is stringent and infeasible when $P_{\mathrm{req}}^{\mathrm{EH}} > 0.12$. This validates the importance of restricting the estimation error with the sensing phase for robust WPT design.
\begin{figure}[t]
\centering
\includegraphics[width=3.3in]{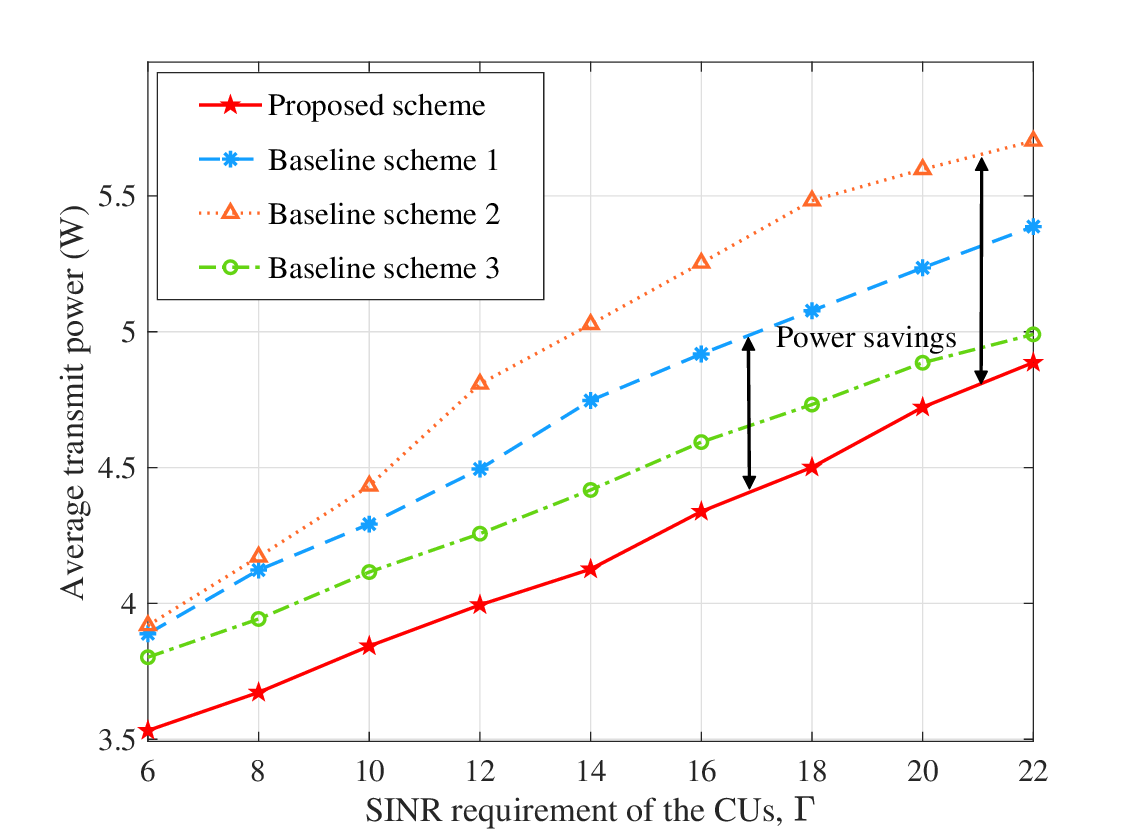}
\caption{Average transmit power versus SINR requirement.}
\label{figure:P_SINR}
\end{figure}
\subsection{Influence of Communication SINR Requirement}
Fig. \ref{figure:P_SINR} demonstrates the power consumption versus the SINR requirement of the CUs $\Gamma$.
It can be observed that the power consumption of the system increases with the SINR requirement. The proposed scheme consumes less power than the baseline schemes. 
The gap between baseline scheme 3 and the proposed scheme becomes smaller when $\Gamma$ is high. This gap exists because the proposed scheme is able to minimize the estimation error through the joint beamforming and AP selection design. However, for high SINR requirements, more power is allocated for the communication service, which reduces the sensing performance improvement that comes from the AP selection. Hence, the gap between baseline scheme 3 and the proposed scheme becomes smaller for high SINR requirements.
\begin{figure}[t]
\centering
\includegraphics[width=3.3in]{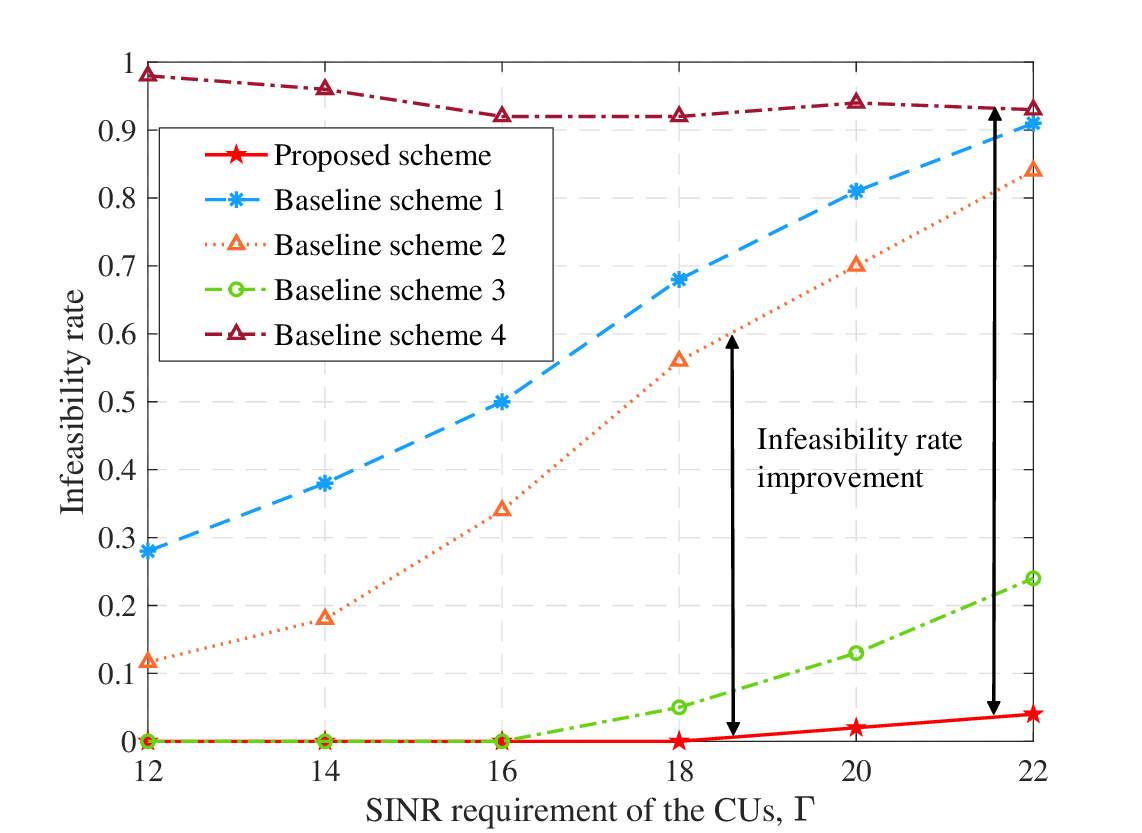}
\caption{Infeasibility rate versus SINR requirement.}
\label{figure:inf_rate}
\end{figure}
\subsection{Infeasiblity Rate}
Fig. \ref{figure:inf_rate} shows the infeasibility rate v.s. the SINR requirement of the CUs. Infeasibility occurs when no solution can simultaneously satisfy all the constraints. 
For $N_{\mathrm{Total}}$ realizations, we solve the problem and count the number of the infeasible solutions, denoted by $N_{\mathrm{Inf}}$. Then the infeasibility rate is calculated by $\frac{N_{\mathrm{Inf}}}{N_{\mathrm{Total}}}$.
As shown in the figure, the infeasibility rate increases with the increase of the SINR requirement $\Gamma$. Compared with the baseline schemes, the proposed scheme can significantly reduce the infeasibility rate. The infeasibility rate gap between the proposed scheme and baseline scheme 1 is rather high and increases quickly with the increasing $\Gamma$. This is because the proposed scheme can adaptively allocate time to the sensing phase to find feasible solutions against the stringent QoS requirements. Note that baseline scheme 4 suffers from a very high infeasibility rate because it allocates as little power as possible to satisfy the harvested energy requirement at the estimated location. As a result, the harvested energy requirement over the location uncertainty area is generally not satisfied.
\section{Conclusion}
This paper studied the robust sensing-assisted SWIPT for mobile EHRs in the networked ISAC system.
A two-phase design was proposed to tackle the location uncertainty due to the movement of the EHR. In particular, each frame is divided into the sensing and WPT phases, where the initial location error is refined in the first phase by collaborative sensing and then utilized for WPT in the second phase.
A two-layer optimization framework was developed to minimize power consumption by jointly optimizing the beamforming policy, AP selection variable, and time-splitting ratio. 
In the inner layer, for a given time-splitting ratio, we solved the robust CRB minimization problem and the robust WPT problem sequentially. Then, in the outer, we determined the time-splitting ratio through line search.
Simulation results validated the effectiveness of the proposed design for power saving. It was shown that there exists an optimal time-splitting ratio and the joint beamforming, AP selection, and time-splitting ratio design makes the proposed algorithm robust against location uncertainty and stringent QoS requirements.
\section*{Appendix}
\subsection*{A. Proof of Proposition 1}
By vectorizing $\mathbf{Y}_q$ in \eqref{stack_Y}, we have
$\tilde{\mathbf{y}}_{q} = \mathrm{vec}(\mathbf{Y}_{q}) = \Tilde{\mathbf{u}}_q + \Tilde{\mathbf{z}}_{q}$, where $\Tilde{\mathbf{u}}_q = \mathrm{vec} \left ( \mathbf{G}_{q} \mathbf{A}_q \mathbf{X} \right ) $, $\Tilde{\mathbf{z}}_q = \mathrm{vec}\left( \mathbf{Z}_q \right)$. $\Tilde{\mathbf{z}}_q \sim \mathcal{CN}(\mathbf{0},\sigma^2_q \mathbf{I}_{NQ \eta L})$. The $(i,j)$-th element of $\mathbf{F}_q$ for estimating $\bm{\chi}$ is given by \cite{kay1993fundamentals}
\begin{equation} \label{fisher_element}
    F^{(q)}_{i,j}\hspace*{-0.5mm}= \frac{2}{\sigma_q^2} \Re\left\{\frac{\partial \Tilde{\mathbf{u}}_q^H}{\partial \bm{\chi}_i} \frac{\partial \Tilde{\mathbf{u}}_q}{\partial \bm{\chi}_j}\right\},
\end{equation}
where $\frac{\partial \Tilde{\mathbf{u}}_q}{\partial \bm{\chi}_i} =   \mathrm{vec} 
\left ( \dot{\mathbf{G}}^{(i)}_{q} \mathbf{A}_q \mathbf{X} \right )$.
$\dot{\mathbf{G}}^{(i)}_{q}$ denotes the derivative of $\mathbf{G}_{q}$ with respect to $\bm{\chi}_i$. Specifically, if $i \in \{ 1,2 \}$, then the expression of $\dot{\mathbf{G}}^{(i)}_{q}$ is given by
\begin{eqnarray}
\dot{\mathbf{G}}^{(i)}_{q} = \left[ \alpha_{1,q} \dot{\mathbf{G}}_{1,q}^{(i)},\cdots, \alpha_{Q,q} \dot{\mathbf{G}}_{Q,q}^{(i)} \right], \hspace{2mm} i \in \{ 1,2 \}, 
\end{eqnarray}
where $\dot{\mathbf{G}}_{q',q}^{(i)} \overset{\triangle}{=} \dot{\mathbf{a}}^{(i)}(\theta_q) \mathbf{a}(\theta_{q'})^H + \mathbf{a}(\theta_q) \dot{\mathbf{a}}^{(i)}(\theta_{q'})^H $ is the derivative of $\mathbf{G}_{q',q}$ with respect to $p_i$. $\dot{\mathbf{a}}^{(i)}(\theta_q)$ is the derivative of the steering vector with respect to $p_i$.
If $ 3 \leq i \leq Q+1$, and the $i$-th term in $\bm{\chi}$ is $\alpha_{Q',q}$, then $\dot{\mathbf{G}}^{(i)}_{q}$ is given by
$\dot{\mathbf{G}}^{(i)}_{q} = \left[ \mathbf{0},\cdots,  \mathbf{G}_{Q',q},\cdots\mathbf{0} \right]$.
If $ Q+2 \leq i \leq 2Q$, and the $i$-th term in $\bm{\chi}$ is $\alpha_{Q',q}$, then $\dot{\mathbf{G}}^{(i)}_{q}$ is given by
$\dot{\mathbf{G}}^{(i)}_{q} = \left[ \mathbf{0},\cdots,  j\mathbf{G}_{Q',q},\cdots\mathbf{0} \right]$.
\par
The derivation of $F^{(q)}_{i,j}$ is given by
\begin{eqnarray} \label{fisher_derivation}
F^{(q)}_{i,j} &\hspace*{-3mm}=\hspace*{-3mm}& \frac{2}{\sigma_q^2} 
\Re \left\{ \mathrm{vec} \left ( \dot{\mathbf{G}}^{(i)}_q \mathbf{A}_q \mathbf{X} \right )^H \left ( 
\mathrm{vec} \left ( \dot{\mathbf{G}}^{(j)}_{q} \mathbf{A}_q \mathbf{X} \right ) \right ) \right\} \notag \\
&\hspace*{-3mm}=\hspace*{-3mm}& \frac{2 \eta L}{\sigma_q^2} \Re \left\{ \mathrm{Tr} \hspace*{-0.5mm}
\left ( \hspace*{-0.5mm}\dot{\mathbf{G}}^{(j)}_{q} \mathbf{A}_q \mathbf{S}\mathbf{A}_q \left [\dot{\mathbf{G}}^{(i)}_q  \right ] ^H  \right ) \right\}.
\end{eqnarray}
\subsection*{B. Proof of Proposition 2}
Denote $\left \{  \mathbf{W}_k^* \right \} $, $\mathbf{R}^*$, $\mathbf{J}^*$, $\overline{\mathbf{W}}_{q,k}^*$, $\overline{\mathbf{R}}_q^*$, $\overline{\mathbf{J}}_q^*$, $a_q^*$, and $(\overline{\epsilon}_{ij}^{(q)})^*$ as the obtained solution of (SDR1).
For a sufficiently large penalty factor $\mu$, the penalty term approaches $0$ and the value of $a_q^*$ takes value from $\{0,1\}$ \cite{9133130}. According to the constraint C3, one of $a_q$, $q \in \mathcal{Q}$, takes the value $0$ and the others are $1$. Suppose that $a_{q_0} = 0$ and $a_q = 1, \forall q \in \mathcal{Q} \setminus\{q_0\}$. Then we can construct the new solution $\left \{  \widetilde{\mathbf{W}}_k^* \right \} $, $\widetilde{\mathbf{R}}^*$, $\widetilde{\mathbf{J}}^*$, $\widetilde{\overline{\mathbf{W}}}_{q,k}^*$, $\widetilde{\overline{\mathbf{R}}}_q^*$, $\widetilde{\overline{\mathbf{J}}}_q^*$, $\widetilde{a}_q^*$, and $(\widetilde{\overline{\epsilon}}_{ij}^{(q)})^*$
, respectively,
\begin{eqnarray} \label{SDR_transformation}
&& \widetilde{\mathbf{W}}_k^* = \frac{\mathbf{W}_k^* \mathbf{A}_{q_0} \mathbf{h}_k \mathbf{h}_k^H \mathbf{A}_{q_0} \mathbf{W}_k^*}{\mathbf{h}_k^H \mathbf{A}_{q_0} \mathbf{W}_k^* \mathbf{A}_{q_0} \mathbf{h}_k}, \label{W} \\ 
&& \widetilde{\mathbf{R}}^* = \underset{k \in \mathcal{K}}{\sum} \mathbf{W}_k^* + \mathbf{R}^* - \underset{k \in \mathcal{K}}{\sum} \widetilde{\mathbf{W}}_k^*, \label{R} \\
&& \widetilde{\overline{\mathbf{W}}}_{q,k}^* = a_q^* \widetilde{\mathbf{W}}_{k}^*, \hspace{2mm} \widetilde{\overline{\mathbf{R}}}_{q}^* = a_q^* \widetilde{\mathbf{R}}^*, \label{WR} \\
&& \widetilde{\mathbf{J}}^* = \mathbf{J}^*, \hspace{1mm} \widetilde{\overline{\mathbf{J}}}^*_q = \overline{\mathbf{J}}_q^*, \hspace{1mm} \widetilde{a}_q^* = a_q^*, \hspace{1mm} (\widetilde{\overline{\epsilon}}_{ij}^{(q)})^* = (\overline{\epsilon}_{ij}^{(q)})^*. \label{sdr_ja}
\end{eqnarray}
According to \eqref{R}, we have
$\underset{k \in \mathcal{K}}{\sum} \widetilde{\mathbf{W}}_k^* +  \widetilde{\mathbf{R}}^* = \underset{k \in \mathcal{K}}{\sum} \mathbf{W}_k^* + \mathbf{R}^*$.
It follows from \eqref{WR}
\begin{eqnarray}
&&\hspace{-5mm} \underset{k \in \mathcal{K}}{\sum} \widetilde{\overline{\mathbf{W}}}_{q,k}^* +  \widetilde{\overline{\mathbf{R}}}_q^* = a_q^*\left( \underset{k \in \mathcal{K}}{\sum} \widetilde{\mathbf{W}}_{q,k}^* +  \widetilde{\mathbf{R}}_q^* \right) \notag \\ 
&&\hspace{-7mm}=  a_q^* \left( \underset{k \in \mathcal{K}}{\sum} \mathbf{W}_k^* + \mathbf{R}^* \right) = \underset{k \in \mathcal{K}}{\sum} \overline{\mathbf{W}}_{q,k}^* + \overline{\mathbf{R}}_q^*, \hspace{1mm} \forall q,
\end{eqnarray}
where the last equality comes from the property of big-M constraints. Hence, constraints $\overline{\mbox{C1}}$ and $\myoverline{\mbox{C10}}$ are satisfied. According to \eqref{WR} and \eqref{sdr_ja}, constraints C3, C4b, C9, $\mbox{C11a-d}$, $\mbox{C12a-d}$ and $\mbox{C13a-d}$ are satisfied.
\par
Next, we show that the SINR constraint $\overline{\mbox{C2}}$ holds for the constructed solution. We rewrite constraint $\overline{\mbox{C2}}$ as
\begin{eqnarray}
& & (1+\frac{1}{\Gamma_k}) \underset{q \in \mathcal{Q}}{\sum}  \mathrm{Tr} \left (  \mathbf{A}_q \mathbf{H}_k \mathbf{A}_q  \left( \mathbf{W}_k - \mathbf{W}_{q,k} \right) \right ) \notag \\
&& \geq  \underset{q \in \mathcal{Q}}{\sum} \underset{k'\in\mathcal{K}}{\sum} \mathrm{Tr} \left (  \mathbf{A}_q \mathbf{H}_k \mathbf{A}_q \left( \mathbf{W}_{k'} - \mathbf{W}_{q,k'} \right) \right ) \notag \\
&& + \underset{q \in \mathcal{Q}}{\sum} \mathrm{Tr} \left ( \mathbf{A}_q \mathbf{H}_k 
\mathbf{A}_q \left( \mathbf{R} - \mathbf{R}_q \right) \right ) + \sigma_k^2, \hspace{1mm}\forall k.
\end{eqnarray}
According to the property of the big-M constraints, we have $\mathbf{W}_{q,k}^* = \mathbf{W}_{k}^*, \forall q \in \mathcal{Q} \setminus \{ q_0\}$, and $\mathbf{W}_{q_0,k}^* = \mathbf{0}$. Then constraint $\overline{\mbox{C2}}$ can be equivalently transformed into
\begin{eqnarray} \label{sdr_sinr}
&& \hspace{-8mm}(1+\frac{1}{\Gamma_k})  \mathrm{Tr} \left (  \mathbf{A}_{q_0} \mathbf{H}_k \mathbf{A}_{q_0} \mathbf{W}_k^* \right ) - \mathrm{Tr} \left ( \mathbf{A}_{q_0} \mathbf{H}_k 
\mathbf{A}_{q_0} \mathbf{R}^*  \right ) \notag \\
&& \hspace{-8mm}- \underset{k'\in\mathcal{K}}{\sum} \mathrm{Tr} \left (  \mathbf{A}_{q_0} \mathbf{H}_k \mathbf{A}_{q_0} \mathbf{W}_{k'}^* \right ) - \sigma_k^2 \geq 0.
\end{eqnarray}
According to \eqref{W}, we can verify that
\begin{eqnarray} \label{sdr_eq}
\mathbf{h}_k^H \mathbf{A}_{q_0} \widetilde{\mathbf{W}}_k^* \mathbf{A}_{q_0} \mathbf{h}_k = \mathbf{h}_k^H \mathbf{A}_{q_0} \mathbf{W}_k^* \mathbf{A}_{q_0} \mathbf{h}_k, \hspace{1mm}\forall k.
\end{eqnarray}
Substituting \eqref{sdr_eq} into \eqref{sdr_sinr}, together with $\underset{k \in \mathcal{K}}{\sum} \widetilde{\mathbf{W}}_k^* +  \widetilde{\mathbf{R}}^* = \underset{k \in \mathcal{K}}{\sum} \mathbf{W}_k^* + \mathbf{R}^*$, we have
\begin{eqnarray}
&& \hspace{-8mm}(1+\frac{1}{\Gamma_k})  \mathrm{Tr} \left (  \mathbf{A}_{q_0} \mathbf{H}_k \mathbf{A}_{q_0} \widetilde{\mathbf{W}}_k^* \right ) - \mathrm{Tr} \left ( \mathbf{A}_{q_0} \mathbf{H}_k 
\mathbf{A}_{q_0} \widetilde{\mathbf{R}}^*  \right ) \notag \\
&&\hspace{-8mm} - \underset{k'\in\mathcal{K}}{\sum} \mathrm{Tr} \left (  \mathbf{A}_{q_0} \mathbf{H}_k \mathbf{A}_{q_0} \widetilde{\mathbf{W}}_{k'}^* \right ) - \sigma_k^2 \geq 0.
\end{eqnarray}
This means that the constructed solution satisfies constraint $\overline{\mbox{C2}}$. Furthermore, according to \eqref{sdr_ja}, the value of the objective function of the constructed solution remains the same after the transformation. Notice that $\mathrm{Rank} \left \{ \widetilde{\mathbf{W}}_k^* \right \} = 1, \forall k.$ holds. This completes the proof.
\bibliographystyle{IEEEtran}
\bibliography{Reference_List}
\end{document}